\RequirePackage{xcolor}

\documentclass{article}

\usepackage{thumbpdf,lmodern}

\usepackage[utf8]{inputenc}
\usepackage{amssymb}
\usepackage{bm}
\usepackage[nomain,symbols,acronym]{glossaries}
\usepackage[automake]{glossaries-extra}
\usepackage{tikz}
\usetikzlibrary{shapes.geometric, arrows,backgrounds,fit,positioning}
\usepackage{comment}
\usepackage{fancyvrb}
\usepackage[authoryear,round]{natbib}
\usepackage{hyperref}
\usepackage{authblk}
\usepackage[margin=4cm]{geometry}
\bibpunct{(}{)}{;}{a}{}{,}

\usepackage{amsmath}

\glsdisablehyper

\newcommand{\varvector}[1]{\ensuremath{\bm{\lowercase{#1}}}}
\newcommand{\varfunction}[1]{\ensuremath{\lowercase{#1}}}
\newcommand{\varFunction}[1]{\ensuremath{\uppercase{#1}}}
\newcommand{\varmatrix}[1]{\ensuremath{\bm{\uppercase{#1}}}}
\newcommand{\randscalar}[1]{\ensuremath{\uppercase{#1}}}
\newcommand{\randvector}[1]{\ensuremath{\bm{\uppercase{#1}}}}
\newcommand{\randfunction}[1]{\ensuremath{\uppercase{#1}}}
\newcommand{\constset}[1]{\ensuremath{\uppercase{#1}}}

\makeglossaries

\setabbreviationstyle[acronym]{long-short}
\glssetcategoryattribute{acronym}{glossdesc}{title}

\newcommand{\placeholder}{\square}
\newcommand{\simpleplaceholder}{{}\cdot{}}

\setglossarypreamble[symbols]{Along this work we will use lowercase letters to refer to scalars or functions. Thus, \(x\) can be a scalar or a function, and it should be clear from context. Sometimes we will use uppercase letters for variables that are limits of a summation or integral, like \(N\). We will use lowercase bold letters to represent vectors, like \varvector{x}, and uppercase bold letters to represent matrices, like \(\varmatrix{x}\). Uppercase letters are also used for sets, random variables and stochastic processes (\(\randscalar{x}\)) and for random vectors (\(\randvector{x}\)). All ambiguities in the notation will be clarified in the text. As a shortcut we will use the notation \(\varfunction{x}(\varvector{t})\) referring to the vector of all evaluations \((\varfunction{x}(t_1), \varfunction{x}(t_2), \ldots)\). Other symbols used in this work are shown below:}

\newglossaryentry{estimated}%
{%
  name={\ensuremath{\hat{\placeholder}}},
  user1={hat},
  description={Sample estimate},
  sort={.},
  type={symbols}
}

\newglossaryentry{nclasses}%
{%
  name={\ensuremath{C}},
  description={Number of classes or clusters in classification and clustering tasks},
  sort={C},
  type={symbols}
}

\newglossaryentry{shift}%
{%
  name={\ensuremath{\delta}},
  description={Shift of the continuous parameter in shift registration},
  sort={d},
  type={symbols}
}

\newglossaryentry{depth_function}%
{%
  name={\ensuremath{\varFunction{D}(\simpleplaceholder)}},
  text={\varFunction{D}},
  description={Depth function},
  sort={D},
  type={symbols}
}

\newglossaryentry{distance_function}%
{%
  name={\ensuremath{\varfunction{d}(\simpleplaceholder)}},
  text={\varfunction{d}},
  description={Distance function},
  sort={d},
  type={symbols}
}

\newglossaryentry{distribution_function}%
{%
  name={\ensuremath{\varFunction{F}_{\randscalar{X}}(\simpleplaceholder)}},
  text={\varFunction{F}},
  description={Distribution function of \(\randscalar{X}\)},
  sort={F},
  type={symbols}
}

\newglossaryentry{basis_function}%
{%
  name={\ensuremath{\phi}},
  description={Element of a basis of functions},
  sort={f},
  type={symbols}
}

\newglossaryentry{warping_function}%
{%
  name={\varfunction{\gamma}},
  description={Warping function},
  sort={g},
  type={symbols}
}

\newglossaryentry{smoothing_parameter}%
{%
  name={\ensuremath{h}},
  description={Smoothing parameter},
  sort={h},
  type={symbols}
}

\newglossaryentry{nbasis}%
{%
  name={\ensuremath{K}},
  description={Number of basis elements used in a basis expansion},
  sort={K},
  type={symbols}
}

\newglossaryentry{kernel_density}%
{%
  name={\ensuremath{\varFunction{k}(\simpleplaceholder)}},
  text={\varFunction{k}},
  description={Kernel density},
  sort={K},
  type={symbols}
}

\newglossaryentry{covariance_function}%
{%
  name={\ensuremath{\varfunction{k}(\simpleplaceholder, \simpleplaceholder)}},
  text={\varfunction{k}},
  description={Covariance function},
  sort={k},
  type={symbols}
}

\newglossaryentry{cov_operator}%
{%
  name={\ensuremath{\mathcal{\varFunction{k}}(\simpleplaceholder)}},
  text={\ensuremath{\mathcal{\varFunction{k}}}},
  description={Covariance operator},
  sort={K},
  type={symbols}
}

\newglossaryentry{L2}%
{%
  name={\ensuremath{L^2}},
  description={Space of (equivalence classes of) square-integrable functions},
  sort={L2},
  type={symbols}
}

\newglossaryentry{sample_points}%
{%
  name={\ensuremath{M}},
  description={Number of observed points per sample},
  sort={M},
  type={symbols}
}

\newglossaryentry{mean_function}%
{%
  name={\ensuremath{\varfunction{\mu}(\simpleplaceholder)}},
  text={\varfunction{\mu}},
  description={Mean function},
  sort={m},
  type={symbols}
}

\newglossaryentry{nsamples}%
{%
  name={\ensuremath{N}},
  description={Number of observations in a sample},
  sort={N},
  type={symbols}
}

\newglossaryentry{real_numbers}%
{%
  name={\ensuremath{\mathbb{R}}},
  description={Set of real numbers},
  sort={R},
  type={symbols}
}

\newglossaryentry{smoothing_matrix}%
{%
  name={\varmatrix{S}},
  description={Smoothing matrix},
  sort={S},
  type={symbols}
}

\newglossaryentry{time_interval}%
{%
  name={\constset{\mathcal{T}}},
  description={Interval where the functions are defined},
  sort={T},
  type={symbols}
}

\newglossaryentry{membership_degree}%
{%
  name={\ensuremath{u}},
  description={Membership degree in fuzzy clustering algorithms},
  sort={u},
  type={symbols}
}

\newglossaryentry{fuzzifier}%
{%
  name={\ensuremath{\omega}},
  description={Fuzzifier used in fuzzy algorithms},
  sort={w},
  type={symbols}
}

\newglossaryentry{smoothing_penalizing_function}%
{%
  name={\ensuremath{\varFunction{\Xi}(\simpleplaceholder)}},
  text={\varFunction{\Xi}},
  description={Penalizing function for linear smoothing},
  sort={X},
  type={symbols}
}

\newacronym{akaike}{AIC}{Akaike’s information criterion}
\newacronym{api}{API}{application programming interface}
\newacronym{bd}{BD}{band depth}
\newacronym{cv}{CV}{cross validation}
\newacronym{fcm}{FCM}{Fuzzy C-Means}
\newacronym{fda}{FDA}{functional data analysis}
\newacronym{fm}{FM-depth}{Fraiman and Muniz depth}
\newacronym{fpca}{FPCA}{functional principal components analysis}
\newacronym{fpe}{FPE}{finite prediction error}
\newacronym{gcv}{GCV}{generalized cross validation}
\newacronym{gm}{GM}{geometric mean}
\newacronym{iqr}{IQR}{interquartile range}
\newacronym{mbd}{MBD}{modified band depth}
\newacronym{mei}{MEI}{modified epigraph index}
\newacronym{mfpca}{MFPCA}{multivariate functional principal components analysis}
\newacronym{mh}{MH}{maxima hunting}
\newacronym{mrmr}{mRMR}{minimum-redundancy-maximum-relevance}
\newacronym{ms}{MS-plot}{magnitude-shape plot}
\newacronym{pca}{PCA}{principal components analysis}
\newacronym{regsse}{REGSSE}{registration sum of squared errors}
\newacronym[shortplural=RKHS]{rkhs}{RKHS}{reproducing kernel Hilbert space}
\newacronym{rmh}{RMH}{recursive maxima hunting}
\newacronym{srsf}{SRSF}{square root slope function}
\newacronym{svm}{SVM}{support vector machine}
\newacronym{ufuncs}{ufuncs}{universal functions}

\newcommand{\class}[1]{\code{#1}}
\newcommand{\fct}[1]{\code{#1()}}
\newcommand{\attribute}[1]{\code{#1}}

\newcommand{\estimated}[1]{\glsxtrfmt{estimated}{#1}}
\DeclareMathOperator{\median}{median}
\newcommand{\R}{\gls{real_numbers}}
\DeclareMathOperator{\tr}{trace}

\providecommand{\keywords}[1]
{
  \small	
  \textbf{\textit{Keywords---}} #1
}

\newcommand{\pkg}[1]{\textbf{#1}}
\newcommand{\proglang}[1]{\textsf{#1}}
\newcommand{\code}[1]{\texttt{#1}}

\DefineVerbatimEnvironment{CodeInput}{Verbatim}{fontshape=sl}
\DefineVerbatimEnvironment{CodeOutput}{Verbatim}{}
\newenvironment{CodeChunk}{}{}

\author[1]{Carlos Ramos-Carreño}
\author[2]{José Luis Torrecilla}
\author[1]{Miguel Carbajo-Berrocal}
\author[1]{Pablo Marcos}
\author[1]{Alberto Suárez}

\date{}

\affil[1]{Department of Computer Science\\
  Escuela Politécnica Superior\\
  Universidad Autónoma de Madrid\\
  Madrid, Spain
}

\affil[2]{Department of Mathematics\\
  Facultad de Ciencias\\
  Universidad Autónoma de Madrid\\
  Madrid, Spain}
  
\affil[ ]{\texttt{\{carlos.ramos, joseluis.torrecilla, alberto.suarez\}@uam.es}}

\title{\pkg{scikit-fda}: A \proglang{Python} Package for Functional Data Analysis}

\begin{document}

\maketitle


\begin{abstract}
The library \pkg{scikit-fda} is a \proglang{Python} package for Functional Data Analysis (FDA). 
It provides a comprehensive set of tools for representation, preprocessing, and exploratory analysis of functional data.
The library is built upon and integrated in \proglang{Python}'s scientific ecosystem. 
In particular, it conforms to the \pkg{scikit-learn} application programming interface so as to take advantage of the functionality for machine learning provided by this package: pipelines, model selection, and hyperparameter tuning, among others. 
The \pkg{scikit-fda} package has been released as free and open-source software under a 3-Clause BSD license and is open to contributions from the FDA community. 
The library's extensive documentation includes step-by-step tutorials and detailed examples of use.
\end{abstract}


\section[Introduction]{Introduction} \label{sec:intro}

\Gls{fda} is the branch of statistics that deals with observations varying over a continuous parameter, such as curves, surfaces, and other types of functions \citep{cuevas_2014_partial, wang++_2016_functional}. 
These types of data appear in many different fields, such as biology \citep{cremona_2019_functional}, demographics \citep{hyndman+shahid_2007_robust}, economics \citep{frois++_2020_forecasting}, energy security \citep{gong++_2021_assessing}, genomics \citep{leng+muller_2006_classification, chen++_2020_human}, medicine \citep{sorensen++_2013_introduction,ferrando++_2020_detecting, horsley++_2021_maternal}, meteorology \citep{beyaztas+yaseen_2019_drought}, oceanography \citep{assuncao++_2020_3d}, traffic control \citep{wagner_2018_functional, hu++_2019_behavioral}, and other areas of application \citep{ullah+finch_2013_applications}.
The functional nature of these data and, in particular, their continuous structure, entails important differences with respect to the classical multivariate statistics. 
These characteristics require the development of specific statistical and computational tools for their analysis.

Due to the growing interest in functional data, several specialized software tools for FDA have emerged in recent years \citep{scheipl_2021_cran}.
One of the main references in the field is the \pkg{fda} package, which is available in \proglang{R} and MATLAB \citep{ramsay++_2020_fda}. 
This general-purpose library provides an implementation of the methods described in \cite{ramsay+silverman_2005_functional} and \cite{ramsay++_2009_functional}. 
It utilizes a basis expansion representation of the functional observations.
Another important reference in the FDA community is the \pkg{fda.usc} package \citep{febrero+oviedo_2012_statistical}, in which the non-parametric approach developed in \cite{ferraty+vieu_2006_nonparametric} is adopted.
One of the contributions of this library is the introduction of a novel structure for the representation of functional data in discrete form, as a collection of measurements at a grid of points.
This \proglang{R} package provides an extensive range of FDA tools, including methods for regression and classification.

A more recent general-purpose \proglang{R} package is \pkg{tidyfun} \citep{scheipl+goldsmith_2020_tidyfun}. 
In this library, a novel data structure (vectors of class \emph{tf}) is introduced to represent functional observations. 
These \emph{tf} vectors can be included as columns in an \proglang{R} \emph{data frame} alongside with other variables. 
Furthermore, they can be manipulated using the tools of the \emph{tidyverse} ecosystem \citep{wickham++_2019_welcome}. 
Another recent contribution is  \pkg{funData} \citep{happ_2020_object}.  
This package provides a representation for discretized univariate and multivariate functional data of arbitrary dimensions based on S4 \proglang{R} classes. 

Finally, a variety of computational tools have been developed to address specific problems in FDA.
Some relevant examples are the packages \pkg{refund} \citep{goldsmith_2019_regression}, which, among others, provides tools for functional regression and principal component analysis (FPCA), \pkg{FDboost} \citep{brockhaus_2018_boosting}, focused on regression problems,
\pkg{funFEM} \citep{bouveyron++_2015_discriminative} and \pkg{funHDDC} \citep{schmutz++_2020_clustering} for functional clustering,
\pkg{fpca} \citep{peng+paul_2011_fpca} and \pkg{fdapace} \citep{carroll_2020_fdapace}, which are mainly devoted to functional principal component analysis (FPCA). 
The \pkg{fdapace} package available in \proglang{R}, and its MATLAB counterpart \pkg{PACE} \citep{yao_2015_pace},
provide methods for both sparsely or densely sampled random trajectories based on FPCA, via the Principal Analysis by Conditional Estimation (PACE) algorithm.
The package \pkg{fdasrvf} \citep{tucker_2020_fdasrvf} contains tools for alignment, elastic registration, PCA, and regression with functional data based on the square-root velocity framework (SRVF) described in \cite{srivastava+klassen_2016_functional}.
This package is available under this name both in \proglang{R} and in \proglang{MATLAB}, as \pkg{fdasrsf} \citep{tucker_2020_fdasrsf_python} in \proglang{Python}, and as \pkg{ElasticFDA} \citep{tucker_2021_elastic} in \proglang{Julia}.
The package \pkg{roahd} \citep{ieva++_2019_roahd} includes a collection of methods for robust analysis of functional data. 
Outlier dectection tools are provided also in \pkg{fdaoutlier} \citep{ojo++_2021_fdaoutlier}.
Visualization tools, including interactive ones, are provided also in the packages \pkg{rainbow} \citep{hyndman+shang_2010_rainbow} and \pkg{refund.shiny} \citep{wrobel++_2016_interactive}.
A recent contribution is the \proglang{R} package \pkg{mlrFDA} described in \cite{pfisterer++_2021_benchmarking}, which gives access and extends the machine learning framework \pkg{mlr} \citep{bisch++_2016_mlr} for the analysis of functional data.

In recent years, the \proglang{Python} language has become more relevant in statistics, data science, and machine learning. 
However, in contrast to the wide variety of alternatives available for FDA in \proglang{R}, the options in \proglang{Python} are much more limited both in number and functionality.
Some \proglang{Python} libraries devoted to FDA are \pkg{fdasrsf}, which has been described earlier, and the recently released \pkg{FDApy} \citep{golovkine_2021_fdapy}, that provides methods for principal component analysis and clustering.

In this context, we present \pkg{scikit-fda}  \citep{ramos++_2022_scikit}, a general-purpose library that makes functional data processing and analysis accessible to the \proglang{Python} community. 
This package offers data structures for the representation of functional data both in discretized form and as a basis expansion, and an extensive set of tools for preprocessing (smoothing, registration and dimensionality reduction), and statistical analysis, including interactive visualization and outlier detection tools. 
In addition, it provides infrastructure to facilitate the application of the machine learning tools of \pkg{scikit-learn} to functional data. 
Comprehensive documentation is supplied that includes installation instructions, tutorials, API reference, and illustrative examples. 

In the short time since its inception, several libraries have been developed using \pkg{scikit-fda} as their foundation \citep{bernard++_2021_curve, consagra++_2022_efficient}.
In addition, it has been employed in a number of recent investigations \citep{fermanian_2020_functional, torrecilla++_2020_optimal, tan++_2020_monash, tan++_2021_time, pegoraro+beraha_2021_fast}.

In what follows, an overview of the functionality provided by the \pkg{scikit-fda} package is given: 
In Section \ref{sec:representation}, the discretized and the basis expansion representations of the functional observations are introduced. 
Interpolation, extrapolation, and derivation tools are presented also in this section. 
In Section~\ref{sec:package}, the functionality of the package is described. 
It includes subsections devoted to preprocessing (smoothing, registration, dimensionality reduction, and variable selection), exploratory analysis (descriptive statistics, depth measures, interactive visualization, and outlier detection), and the integration with the machine learning tools of \pkg{scikit-learn}. 
Examples of code that illustrate \pkg{scikit-fda}'s functionality are provided throughout the article. 
The tools and processes implemented to ensure the quality of the code are described Section \ref{sec:documentation}. 
This section includes also an overview of the project's extensive documentation. 

\section[Representation of functional data in scikit-fda]{Representation of functional data in \pkg{scikit-fda}} \label{sec:representation}

The \pkg{scikit-fda} package provides tools for the representation of functional observations of the form \(x:\gls{time_interval} \subseteq \R^p \to \R^q\), with \(p \ge 1\), and \(q\ge 1\).
The parameter \(p\) is the dimension of the domain of the functions (\(p = 1\) for curves, \(p = 2\) for surfaces, etc.). 
The parameter \(q\) is the dimension of the codomain; that is, the number of output coordinates for vector-valued functions. 
For instance, a grayscale two-dimensional image can be treated as a functional datum with \(p=2\), for the location of the pixels in the image, and \(q=1\), for the intensity at each pixel.
A color image consisting of three channels (e.g., red, green, and blue) would have \(p=2\) and \(q=3\).
The values of \(p\) and \(q\) are the same for all the observations in a functional dataset.
For the sake of clarity, we focus on the case of real-valued univariate functions (that is,  \(p=q=1\))  
for which most of the functionality described in this article is implemented.
We will further assume that the functions are defined on a compact interval in the real line.
In higher dimensions, the domain is assumed to be rectangular.
Typically, the continuous parameter on which the functions depend, $t\in \gls{time_interval}$, is assumed to be time.

\begin{figure}[t]
  \centering
  \includegraphics[width=0.5\linewidth]{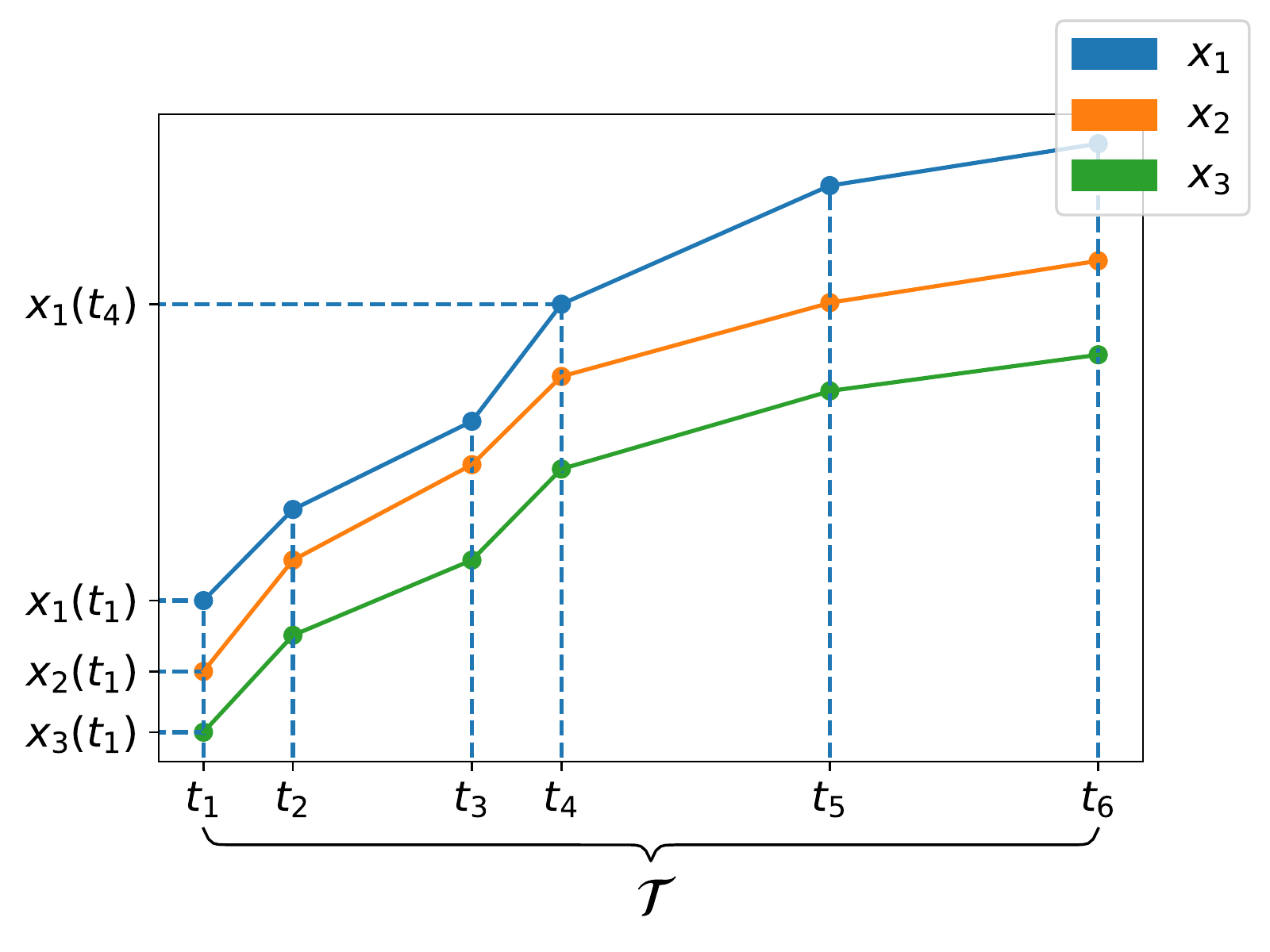}
  \caption{Functional observations in discretized form. The quantity $x_n(t_j)$ represents the value of the $n$-th trajectory at $t_j$. }
  \label{fig:notation}
\end{figure}

A functional dataset consists of a collection of \(\gls{nsamples}\) observations \(\left\{\varfunction{x}_i(t), t\in \gls{time_interval} \right\}_{i=1}^{\gls{nsamples}}\),
where \(\varfunction{x}_i(t)\) is the \(i\)-th observation in the sample.
Each observation can be represented either in discretized form, or as a basis expansion.
In the former representation, a functional observation consists of a set of measurements at a grid of points \(\varvector{t} = (t_1, \ldots, t_{\gls{sample_points}})\in \gls{time_interval}^M\), which is common for all observations. 
This grid need not be regularly spaced. 
The \(i\)-th observation in the sample is represented by the vector \(\left\{\varfunction{x}_i(\varvector{t}) \right\}_{i=1}^{\gls{nsamples}}\), where \(\varfunction{x}_i(\varvector{t}) = (\varfunction{x}_i(t_1), \ldots, \varfunction{x}_i(t_{\gls{sample_points}}))\).
The discretization grid is assumed to be sufficiently fine so that the functional character of the data is apparent \citep{ramsay+silverman_2005_functional,ferraty+vieu_2006_nonparametric,hsing+eubank_2015_theoretical}.
As an illustration of this representation, three example trajectories measured at a grid of irregularly-spaced points are displayed in Figure~\ref{fig:notation}. 

Alternatively, a functional observation can be represented as an expansion in a functional basis \(\{\gls{basis_function}_i(t), t\in \gls{time_interval} \}_{i\ge 1}\)
\begin{equation}\label{eq:functional_basis_expansion}
	\varfunction{x}(t) = \sum_{i \ge 1} c_{i}\gls{basis_function}_i(t),
\end{equation} 
where $ \left\{c_{i}\right\}_{i \ge 1}$ are the coefficients of the expansion.   

The package \pkg{scikit-fda} provides data structures for both types of representation: the class \class{FDataGrid} for discretized data, and the class \class{FDataBasis} for expansions in a functional basis. 
They are derived from the abstract class \class{FData}, which provides common properties and methods. 
In what follows, these classes are described in detail. 

\subsection[The class FData]{The class \class{FData}}

In the class \class{FData}, the attributes and methods shared by the discretized and the basis expansion representations of the functional dataset are collected. 
Thus, it provides a common interface for both \class{FDataGrid} and \class{FDataBasis} objects. 
Specifically, objects of the \class{FData} class have the following attributes:

\begin{itemize}
\item \attribute{dataset\_name}: Name of the functional dataset.
  \item \attribute{n\_samples}: Size (number of functional observations) of the dataset. 
  \item \attribute{dim\_domain}: Dimension of the domain in which the functions are defined (\(p \ge 1\)).
  \item \attribute{argument\_names}: Names of each of the \(p\) arguments of the function (domain dimensions). 
  \item \attribute{domain\_range}: Limits of the intervals for each of the domain arguments. They are used as the default ranges for plotting and numerical integration.
  \item \attribute{dim\_codomain}: Dimension of the codomain \(q\) (output). For scalar functions, \(q=1\). Values \(q > 1\) correspond to vector-valued functions.
  \item \attribute{coordinate\_names}: Names of the \(q\) codomain coordinates.
  \item \attribute{extrapolation}: Default extrapolation strategy; for instance, constant or periodic. See Section~\ref{sec:additionalrepr} for details.
\end{itemize}

As an illustration of this data structure, consider the case of observations that are bidimensional RGB images. 
For this functional dataset, \attribute{dim\_domain} would be 2, corresponding to the two dimensions of the image. 
The names of these dimensions (e.g., ``x'' and ``y'', or ``horizontal'' and ``vertical'') would be stored in the attribute \attribute{argument\_names}.
The attribute \attribute{dim\_codomain} would be 3, corresponding to the three color channels.
The names of these channels (``R'', ``G'', and ``B'') would be stored in the attribute \attribute{coordinate\_names}.

The class \class{FData} provides also methods that are common to both representations; for instance,  methods for evaluation, addition, multiplication by a scalar, and plotting. 
Since \class{FData} is abstract, it is not possible to directly instantiate an object of this class.
Instead, objects of one of its subclasses, \class{FDataGrid} or \class{FDataBasis}, need to be created. 

\subsection[Discretized representation: The class FDataGrid]{Discretized representation: The class \class{FDataGrid}}\label{sub:FDataGrid}

Functional data are often the result of monitoring a continuous process at a discrete set of points.
For the general case, \(x: \R^p \to \R^q\), with \(p, q \ge 1\), we assume that the discretization grid in the \(j\)-th dimension is \(\varvector{t}_j = (t_{j1}, \ldots, t_{j\gls{sample_points}_j})\), with \(j=1, \ldots, p\). 
In the \pkg{scikit-fda} library, the points in the grid need not be regularly spaced.
The grid has to be the same for all observations in the functional dataset.
The dataset  \(\left\{ x_i: \R^p \to \R^q \right\}_{i=1}^{\gls{nsamples}}\) is represented as an object of the class \class{FDataGrid}.
The values of the functional observations are stored in the tensor \(\{\varfunction{x}_i(\varvector{t}) = \varfunction{x}_i(\varvector{t}_1 \times \ldots \times \varvector{t}_p)\}_{i=1}^{\gls{nsamples}}\) of dimension \(\gls{nsamples} \times \gls{sample_points}_1 \times \cdots \times \gls{sample_points}_p\times q\).
Here \(\varvector{t}_1 \times \ldots \times \varvector{t}_p\) is the grid of points obtained as the cartesian product of \(\varvector{t}_1, \ldots, \varvector{t}_p\).
In the case \(p = q = 1\), the discretized sample is simply an \(\gls{nsamples} \times \gls{sample_points}\) matrix \(\{\varfunction{x}_i(\varvector{t}) = (\varfunction{x}_i(t_1), \ldots, \varfunction{x}_i(t_{\gls{sample_points}}))\}_{i=1}^{\gls{nsamples}}\). 

In addition to those inherited from \class{FData}, objects of this class have the following attributes:
\begin{itemize}
   \item \attribute{grid\_points}: Sequence of discretization grids, one for each of the domain dimensions \(\left(\varvector{t}_1, \varvector{t}_2, \ldots, \varvector{t}_p \right)\). 
   The values of the functions are specified at the Cartesian product of the 1-D arrays in the sequence. 
  \item \attribute{data\_matrix}: \pkg{NumPy} array of dimensions \(\gls{nsamples} \times \gls{sample_points}_1 \times \cdots \times \gls{sample_points}_p\times q\) in which the values of the \(\gls{nsamples}\) functional observations are stored. 
\item \attribute{interpolation}: Default interpolation strategy for locations within the rectangular discretization grid. See Section~\ref{sec:additionalrepr} for details. 
\end{itemize}

Since the attribute \attribute{data\_matrix} is a \pkg{NumPy} array, it is possible to carry out pointwise operations, such as powers, exponentials, logarithms, and trigonometric functions by directly applying the
corresponding \pkg{NumPy} functions \citep{harris++_2020_array}.

As an illustration,  in the following code a \class{FDataGrid} object is created with three functional observations measured at grid points \(\varvector{t}=(0, 0.1, 0.3, 0.4, 0.7, 1)\).
These data are depicted in Figure~\ref{fig:notation}.

\begin{CodeInput}
import skfda

grid_points = [0.0, 0.1, 0.3, 0.4, 0.7, 1.0]
data_matrix = [
    [109.5, 115.8, 121.9, 130.0, 138.2, 141.1],
    [104.6, 112.3, 118.9, 125.0, 130.1, 133.0],
    [100.4, 107.1, 112.3, 118.6, 124.0, 126.5],
]

fd = skfda.FDataGrid(
    data_matrix=data_matrix,
    grid_points=grid_points,
)
\end{CodeInput}

In the previous example, the discretization points and the values of the functions are specified manually. 
The values of \attribute{grid\_points} and \attribute{data\_matrix} can be imported from data files in standard formats, such as CSV, XLSX, ARFF, MATLAB files, with the help of the corresponding functions from \pkg{NumPy} \citep{harris++_2020_array}, \pkg{SciPy} \citep{virtanen++_2020_scipy}, \pkg{pandas} \citep{pandas_2020_pandas}, and similar packages.
An example of how to import data from a CSV file is provided in \url{https://fda.readthedocs.io/importing_data}.

\subsection[Basis expansion representation: The class FDataBasis]{Basis expansion representation: The class \class{FDataBasis}}

{
\newcommand{\samplefunction}{\varfunction{x}(t)}
\newcommand{\basissumbody}{c_{i}\gls{basis_function}_i(t)}

Assume that the functional observations in the daset considered  belong to $ \mathcal{F}$, a separable Hilbert space; for instance, \(\gls{L2}\). 
Under such assumption there exists a countable basis \(\{\gls{basis_function}_i(t)\}_{i\geq 1}\) that is complete, so that any  $\samplefunction \in \mathcal{F}$ can be expressed as
\begin{equation}\label{eq:functional_basis_general_expansion}
	\samplefunction = \sum_{i \ge 1}\basissumbody,
\end{equation}
where \(\left\{ c_{i} \right\}_{i \ge 1}\), are the coefficients of basis expansion. 
The \pkg{scikit-fda} library provides support for such a representation in the constant, monomial, B-spline, and Fourier bases.
In addition, other types of bases can be implemented by the user.
An example showing how to define new bases is available in \url{https://fda.readthedocs.io/create_new_bases}.
The choice of basis should be made taking into consideration the characteristics of the data at hand.
For instance, the Fourier basis is well-suited to representing periodic functions.
For non-periodic data, a representation in the B-splines basis is probably more appropriate.
The monomial basis is useful to represent polynomials.
Monomials are also the building blocks of the Maclaurin series. 
Therefore, the monomial basis can be used to represent local approximations of analytic functions.
The first five elements of the different bases provided in \pkg{scikit-fda} are shown in Figure \ref{fig:basis}.
\begin{figure}
  \includegraphics[width=\linewidth]{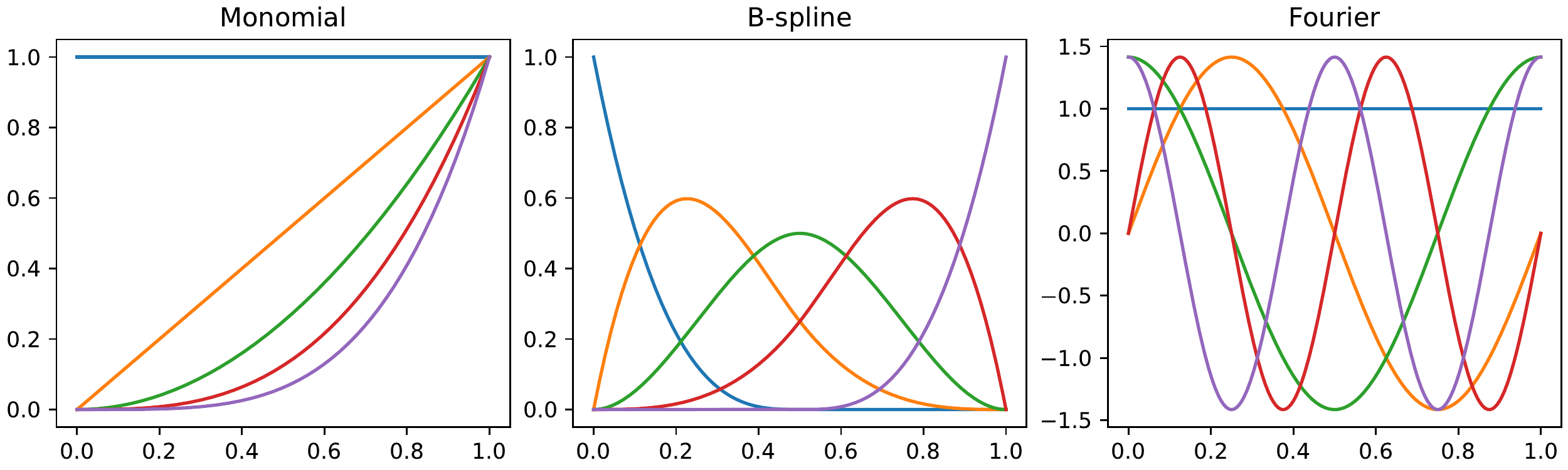}
  \caption[Illustration of available bases]{First five elements of the bases available in \pkg{scikit-fda}: Monomial (left), B-splines (center), and Fourier (right).}
  \label{fig:basis}
\end{figure}

In general, these types of representation are infinite-dimensional. 
In practice, the expansion is truncated to the first \(\gls{nbasis}\) terms
\begin{equation}\label{eq:truncated_basis}
	\samplefunction \approx \sum_{i=1}^{\gls{nbasis}} \basissumbody.
\end{equation}
Truncation often results in the smoothing of the original functional observations. This smoothing effect can be beneficial for the representation of noisy data (see Section~\ref{sec:smoothing}).
} 

In \pkg{scikit-fda}, the class \class{FDataBasis} is used to represent functional data as a finite basis expansion.
In addition to those inherited from \class{FData}, objects of this class have the following attributes:
\begin{itemize}
  \item \attribute{basis}: The basis for the representation of the functional observations.
  It is an object of the class \class{Basis}, one of whose attributes, \attribute{n\_basis} is the number of elements of the basis considered.
  The bases available are \class{Constant}, \class{Monomial}, \class{BSpline} and \class{Fourier}.
  
  \item \attribute{coefficients}: The \(\gls{nsamples}\times \gls{nbasis}\) matrix that contains the coefficients of the basis expansion. 
\end{itemize}

The following code is used to illustrate \pkg{scikit-fda}'s support for this type of representation.

\begin{CodeInput}
import skfda
from skfda.representation.basis import Fourier, BSpline

X, y = skfda.datasets.fetch_phoneme(return_X_y=True)

X.to_basis(BSpline(n_basis=5))
X.to_basis(Fourier(n_basis=5))
\end{CodeInput}

\begin{figure}
  \includegraphics[width=\linewidth]{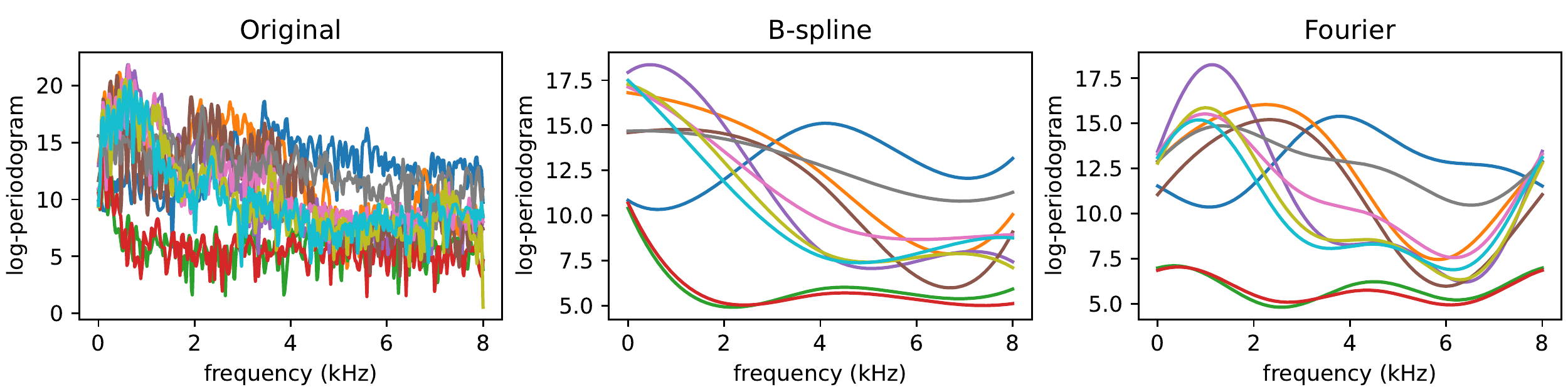}
  \caption[Representation as a basis expansion]{Different representations of the first ten trajectories of the \emph{Phoneme} dataset. From left to right: original trajectories, B-spline, and Fourier basis representation. In both cases, $5$ basis functions are considered}
  \label{fig:fdatabasis}
\end{figure}

In this code, the \emph{Phoneme} dataset \citep{hastie++_2009_elements} is used.
The curves in this dataset correspond to log-periodograms of the time series of utterances of five different phonemes by different speakers. 
The functional observations are transformed from their original discretized representation into different basis expansions.
A sample of the transformed trajectories, together with the original ones, is displayed in Figure~\ref{fig:fdatabasis}. 
In these plots, the smoothing effect of the transformation to the basis representations is apparent.

\subsection{Interpolation and extrapolation} \label{sec:additionalrepr} 

The \pkg{scikit-fda} package provides a variety of interpolation methods for functional data in discretized form. By default, linear interpolation is performed. 
Other types of interpolation can be specified in the attribute \attribute{interpolation} of the \class{FDataGrid} object.
Specifically, support for spline interpolation is provided by the class \class{SplineInterpolation}. 
It is also possible to employ other interpolation and extrapolation strategies defined by the user.
An example showing how to define such custom strategies is available at \url{https://fda.readthedocs.io/create_new_interpolation}.

Different extrapolation strategies for \class{FDataGrid} and \class{FDataBasis} objects are available in \pkg{scikit-fda}. In particular, it is possible to specify a constant value (for instance, the value of the function at one of the limits of the domain), or to assume a periodic structure. In the case of  \class{FDataBasis} objects, one can also directly evaluate the basis expansion outside the domain. Finally, as in interpolation, user-defined extrapolation strategies can be utilized.

\subsection{Derivatives} \label{sec:derivatives}

The computation of derivatives is of particular importance in functional data analysis. 
For instance, derivatives can reveal significant information that is not apparent in the original curves. 
Furthermore, the norm of a derivative is a natural measure of the function's roughness \citep{ramsay+silverman_2005_functional}. 
For this reason, they are often employed to define penalties for regularization.
In the \pkg{scikit-fda} package, the method \fct{derivative} can be used to perform this operation for both \class{FDataGrid} and  \class{FDataBasis} objects.
In the case of \class{FDataGrid} objects, derivatives are approximated using finite differences.
For \class{FDataBasis} objects, they are computed exactly in terms of the derivatives of the basis functions.
Therefore, if a new type of basis is designed, it is necessary to implement the derivatives of the basis functions in the corresponding class.

\subsection{Regularization} \label{sec:regularization}

Regularization methods consist in favoring simpler models to improve the quality and robustness of the solutions of an optimization problem.
In FDA, regularization is used to obtain smooth functional approximations to noisy discrete data, for registration, and for principal component analysis, among others.
For the purpose of regularization, the complexity of a function can be quantified in terms of its norm, or of a linear transformation thereof (e.g., the function derivatives).  
A penalty term proportional to this measure of complexity is then added to the cost function to be minimized. 
The package \pkg{scikit-fda} provides the necessary infrastructure to implement regularization based on the \(\gls{L2}\)-norm of the function in class \class{L2Regularization}.
Alternatively,  a linear operator can be passed as a parameter to the constructor of \class{L2Regularization} objects. 
Some common linear operators are readily available in \pkg{scikit-fda}'s operators module for that purpose.
The following code illustrates this type of regularization to obtain smooth representations of a set of functions in the basis of B-splines.
In this example, the $L^2$ norm of second derivatives of the trajectories is used to penalize their curvature.

\begin{CodeInput}
import skfda

X, y = skfda.datasets.fetch_phoneme(return_X_y=True)
X = X.coordinates[0]

basis = skfda.representation.basis.BSpline(
    domain_range=X.domain_range,
    n_basis=40,
)

regularization = skfda.misc.regularization.L2Regularization(
    skfda.misc.operators.LinearDifferentialOperator(2),
    regularization_parameter=1,
)

smoother = skfda.preprocessing.smoothing.BasisSmoother(
    basis=basis,
    regularization=regularization,
    return_basis=True,
)

X_basis = smoother.fit_transform(X)
\end{CodeInput}

\begin{figure}
  \includegraphics[width=\linewidth]{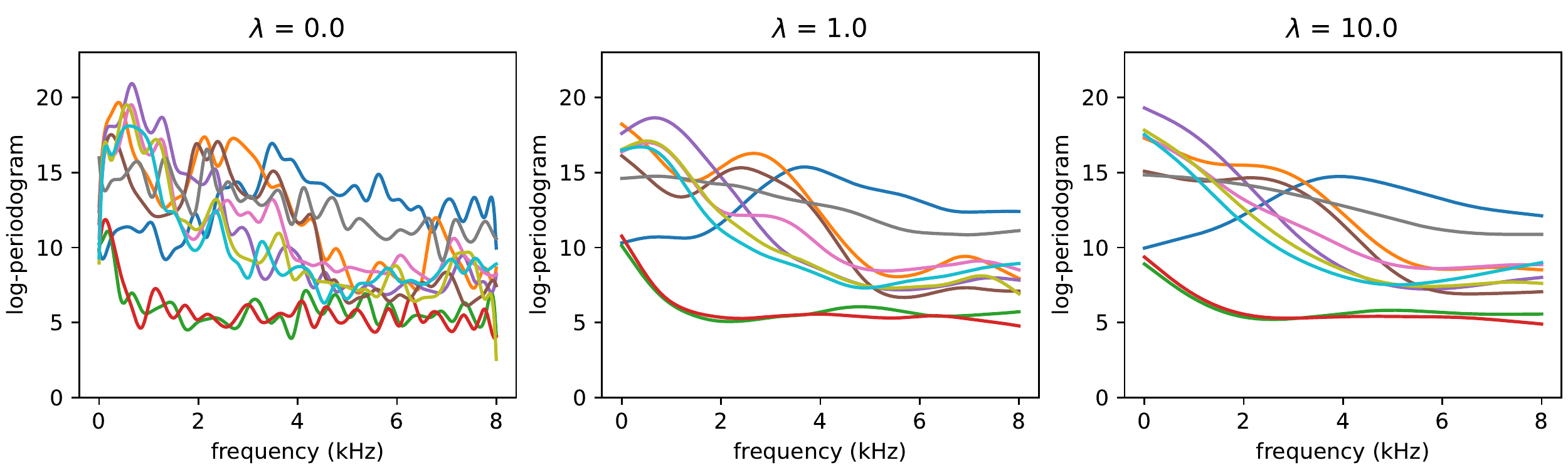}
  \caption[Regularized representation in basis]{Smoothed representation of the first ten trajectories of the \emph{Phoneme} dataset in a B-spline basis with \(40\) basis functions for different values for the regularization parameter \(\lambda \ge 0\);
  from left to right: \(\lambda = 0\) (no regularization), \(\lambda = 1\), and \(\lambda = 10\).}
  \label{fig:regularization}
\end{figure}

The effect of this type of smoothing is illustrated in Figure~\ref{fig:regularization}.
In this figure, \(10\) trajectories of the \emph{Phoneme} dataset are represented in a B-spline basis composed of \(40\) basis functions for different values for the regularization parameter \(\lambda\); from left to right: \(\lambda = 0\) (no regularization), \(\lambda = 1\), and \(\lambda = 10\).
As shown in the left plot of this figure, when the number of basis functions is large, the non-regularized trajectories represented in this basis still exhibit significant fluctuations.
Progressively smoother basis representations are obtained as the \emph{regularization parameter} \(\lambda > 0\) increases.

\section[Functionality of scikit-fda]{Functionality of \pkg{scikit-fda}} \label{sec:package}

In this section, an overview of the utilities for functional data analysis provided by the \pkg{scikit-fda} package is given. 
The first step in the analysis is to generate functional datasets or to retrieve them from external sources. 
In the \pkg{scikit-fda} package it is possible to generate synthetic data, to simulate random trajectories from stochastic processes, and to load data from files in standard formats and from repositories of real-word datasets.
The library provides also an extensive set of tools for the analysis of functional data both in discretized and basis expansion forms.
In particular, it offers methods for exploratory analysis, smoothing, registration, dimensionality reduction, the computation of functional depths, outlier detection, and interactive visualization, among others.
An important feature of \pkg{scikit-fda} is the integration with the extensive collection of \pkg{scikit-learn}'s tools for machine learning, including data preprocessing, training, testing, and hyperparameter selection. Specifically, the methods provided are designed so that they can be utilized in \pkg{scikit-learn} \emph{pipelines}.
In what follows these functionalities will be described in detail.

\subsection{Generation of synthetic data}

A variety of methods for the generation of functional data, either from some simple models or sampled from stochastic processes, are available in \pkg{scikit-fda}.
In particular, the function \fct{make\_multimodal\_samples} can be used to generate functions with several maxima.
This is used in the synthetic registration example illustrated in Figure~\ref{fig:elastic_registration}.
The \fct{make\_gaussian\_process} function can be used to simulate trajectories sampled from  a Gaussian process with a specified mean and covariance function.
Several commonly employed covariance functions, such as Brownian, exponential, radial basis function (RBF), Matérn, and polynomial kernels are supplied in the package. Additional types of covariance functions can be defined by the user.

\begin{figure}
  \centering
  \includegraphics[width=\linewidth]{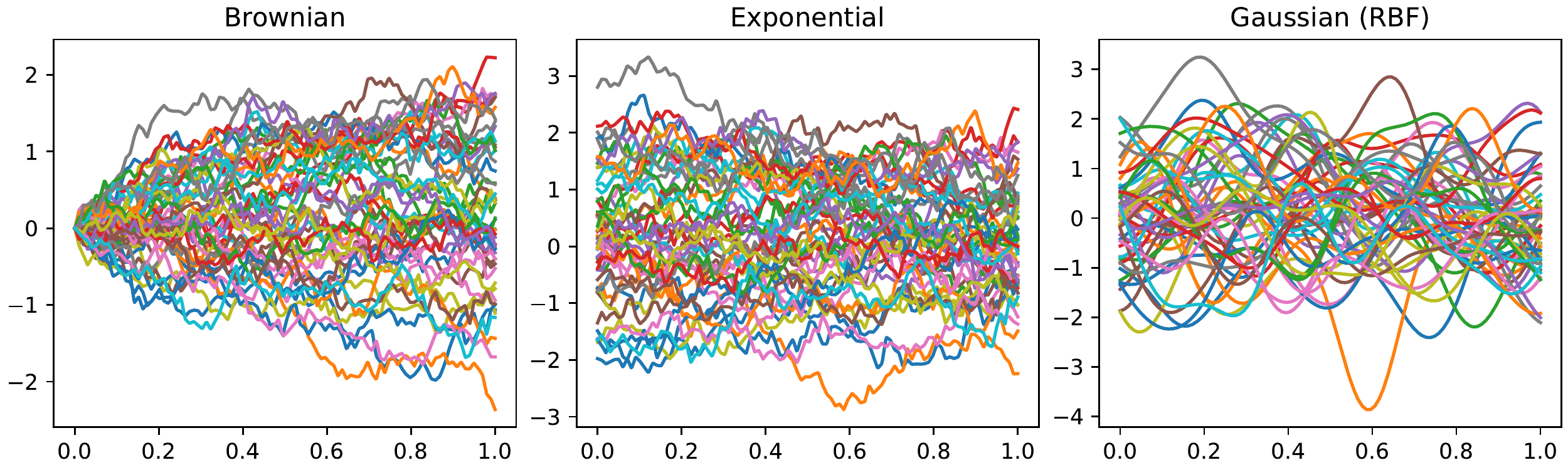}
  \caption[Generation of Gaussian trajectories]{Trajectories sampled from Gaussian processes with different covariance functions. From left to right, standard Brownian, Exponential ($l=1$) and RBF ($l=0.1$).}
  \label{fig:gaussian}
\end{figure}

The following code illustrates the generation of \(50\) trajectories from standard Brownian Motion, an Ornstein-Uhlenbeck process
(a Gaussian process with an exponential covariance function), and a Gaussian process with an RBF covariance function.
A regular grid of \(100\) equally spaced points in \(\gls{time_interval}=[0, 1]\) is employed for the discretized representation. 
The simulated trajectories are displayed in Figure~\ref{fig:gaussian}.

\begin{CodeInput}
import skfda
from skfda.misc.covariances import Brownian, Gaussian, Exponential

cov_dict = {
    "Brownian": Brownian(variance=1),
    "Exponential": Exponential(variance=1, length_scale=1),
    "Gaussian (RBF)": Gaussian(length_scale=0.1),
}

for i, (name, cov) in enumerate(cov_dict.items()):

    fd = skfda.datasets.make_gaussian_process(
        n_samples=50,
        n_features=100,
        mean=0,
        cov=cov,
        random_state=0,
    )

    fd.plot()
\end{CodeInput}

\subsection{Real-world data}

The package \pkg{scikit-fda} provides tools to retrieve functional datasets from other libraries and public access repositories.
The datasets themselves are retrieved using the package \pkg{scikit-datasets} \citep{diaz+ramos_2022_scikit}.  
The data are downloaded only once and cached on disk, so as to reduce network traffic and make it possible to work with the data offline.
They are then converted to the \class{FData} format for further processing and analysis.
For instance, the function \fct{fetch\_cran} can be used to retrieve datasets from \proglang{R} packages in the CRAN repository. 
Datasets from the  \emph{UCR \& UEA Time Series Classification Archive} \citep{dau++_2019_ucr,bagnall++_2018_uea, bagnall++_uea}
can be accessed making use of the function \fct{fetch\_ucr}. 
Moreover, there exist specific functions to import some widely-used datasets such as \fct{fetch\_growth} for the \emph{Berkeley Growth Study} dataset \citep{tuddenham_1954_physical} and \fct{fetch\_weather} for the \emph{Canadian Weather} dataset \citep{ramsay+silverman_2005_functional}.

The following code illustrates the functionality described for three well-known datasets: 
\emph{GunPoint} from the UCR repository, \emph{Canadian Weather}, and the \emph{Berkeley Growth Study}.
\pkg{scikit-fda}'s functions are used to load the data and plot some of the datasets' trajectories.
Note that the \emph{Canadian Weather} dataset has two codomain dimensions (\(q=2\)): temperature and precipitation. 
In this example, the first is selected using the \attribute{coordinates} property of an \class{FData} object. 

\begin{CodeInput}
import skfda

dataset = skfda.datasets.fetch_ucr("GunPoint")
dataset["data"].plot(group=dataset["target"])

X, _ = skfda.datasets.fetch_weather(return_X_y=True)
X.coordinates[0].plot()

X, y = skfda.datasets.fetch_growth(return_X_y=True)
X.plot(group=y)
\end{CodeInput}

\begin{figure}
  \centering
  \includegraphics[width=\linewidth]{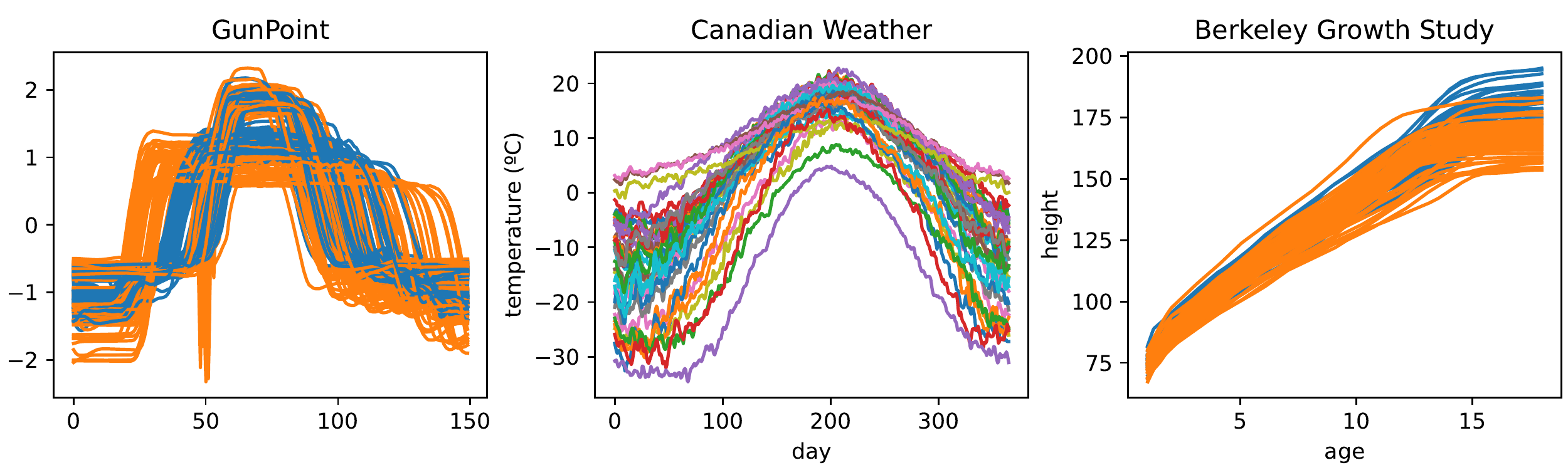}
  \caption[Illustration of real datasets]{Real datasets fetched with \pkg{scikit-fda}. From left to right: \emph{GunPoint}, \emph{Canadian Weather}, and the \emph{Berkeley Growth Study}.}
  \label{fig:fetch_data}
\end{figure}

\subsection{Preprocessing}

Functional observations often need to be subject to some form of processing to facilitate ulterior manipulation. 
To this end, the \pkg{scikit-fda} package provides utilities for smoothing, registration, and dimensionality reduction. 
In what follows, these utilities are described in detail.

\subsubsection{Smoothing}\label{sec:smoothing}

Smoothing consists in replacing the original functional observation \(\varfunction{x}(t)\) with a smoothed version \(\estimated{\varfunction{x}}(t)\). 
This replacement yields a more manageable, possibly more faithful, representation of the underlying process. 
In particular, smoothing can be used to recover the signal component from noisy measurements.
Furthermore, smoothed approximations with continuous derivatives can be used for regularization \citep{wang++_2016_functional}.
The methods of \pkg{scikit-fda}'s classes \class{BasisSmoother}, \class{NadarayaWatsonSmoother}, \class{LocalLinearRegressionSmoother}, and \class{KNeighborsSmoother} can be employed to this end. 
The package provides also utilities to determine an appropriate degree of smoothing using some form of statistical validation.

As discussed in Section~\ref{sec:representation}, the approximation of a function by a truncated basis expansion, as in Equation~\ref{eq:truncated_basis}, is a form of smoothing. 
The coefficients of the finite basis expansion can be estimated by least squares \citep{ramsay+silverman_2005_functional}.
This kind of smoothing is performed in \pkg{scikit-fda} by the class \class{BasisSmoother}. 
Further smoothing can be achieved by a regularization approach based on roughness penalties, as described in Section~\ref{sec:regularization} \citep{green+silverman_1993_nonparametric,ramsay+silverman_2005_functional}. 

Smoothing can be achieved also by performing a linear transformation of the original functional observations
\begin{equation} \label{linear_transformation_smoothing}
	\estimated{\varfunction{x}}(t) = \int_{\gls{time_interval}} \varfunction{s}_{t}(\tau) \varfunction{x}(\tau) d\tau. 
\end{equation}
The weighting function \(\varfunction{s}_{t}(\tau)\) quantifies the contribution of the value of the function at \(\tau\) to the smoothed value at \(t\). This weighting function should be localized, so that the values of the function at points close to \(t\) contribute more to the average.

For functional data in discretized form, Equation~\ref{linear_transformation_smoothing} can be expressed as a matrix transformation
\begin{equation}\label{eq:linear_smoothing_discretized}
	\estimated{\varvector{x}} = \gls{smoothing_matrix} \varvector{x},
\end{equation}
where \(\varvector{x} = \varfunction{x}(\varvector{t})\) is the vector of values of the function at the discretization points \(\varvector{t} = (t_1, \ldots, t_{\gls{sample_points}})\), 
the vector \(\estimated{\varvector{x}}\) consists of the smoothed function values at a grid of points, which can be different from the original ones, and $\gls{smoothing_matrix}$ is the smoothing matrix. 
By default, the grid at which the smoothed values are computed is \(\varvector{t}\), the set of sampling points.
The smoothing matrix \(\gls{smoothing_matrix}\) is sometimes referred to as the ``hat'' matrix, a name borrowed from regression analysis because it transforms the dependent variable vector \(\varvector{x}\) into its fitted version \(\estimated{\varvector{x}}\) \citep{ramsay+silverman_2005_functional}.

Smoothing with local weights can be implemented using kernels  \citep{wasserman_2006_all}.
\pkg{scikit-fda} provides three smoothers of this type: the Nadaraya-Watson (\class{NadarayaWatsonSmoother}), the local linear regression (\class{LocalLinearRegressionSmoother}), and the \(k\) nearest neighbors (\class{KNeighborsSmoother}) smoothers.
As an illustration, for the the Nadaraya-Watson smoother, the hat matrix is
\begin{equation}\label{eq:nadaraya_watson}
	S_{ij}(\gls{smoothing_parameter}) = \frac{\gls{kernel_density}\left(\frac{t_{i}-t_{j}}{\gls{smoothing_parameter}}\right)}{\sum_{m=1}^{\gls{sample_points}}\gls{kernel_density}\left(\frac{t_{i}-t_{m}}{\gls{smoothing_parameter}}\right)}, \quad 1 \leq i, j \leq \gls{sample_points},
\end{equation}
where \(\gls{kernel_density}\) is the kernel function and \(\gls{smoothing_parameter}\) is the parameter that controls the degree of smoothing.
Commonly used kernel functions, such as Gaussian, uniform, and Epanechnikov, are available in \pkg{scikit-fda}. 
It is also possible to employ user-defined kernels for this type of smoothing.
 
In these methods, the value of the smoothing parameter needs to be carefully adjusted to avoid under- or oversmoothing. 
In \pkg{scikit-fda} this value can be determined by statistical validation with the help of the class \class{SmoothingParameterSearch}.
Particular scoring criteria, such as the leave-one-out cross-validation (\class{LinearSmootherLeaveOneOutScorer}) or, alternatively, generalized cross-validation  (\class{LinearSmootherGeneralizedCVScorer}) are provided to guide the search.
The criterion that is maximized in leave-one-out cross validation is
\begin{equation}\label{eq:loo_smothing_criterion}
CV_{loo}(\gls{smoothing_parameter})=\frac{1}{\gls{sample_points}} \sum_{m=1}^{\gls{sample_points}} \left(\frac{x(t_m) - \estimated{x}(t_m; \gls{smoothing_parameter})}{1 - S_{mm}(\gls{smoothing_parameter})}\right)^2,
\end{equation}
where \(\gls{smoothing_parameter}\) is the smoothing parameter and \( \estimated{x}(t_m; \gls{smoothing_parameter}) \) is the smoothed value.

The \gls{gcv} criterion is
\begin{equation}\label{eq:gcv_smothing_criterion}
\gls{gcv}(\gls{smoothing_parameter})=\gls{smoothing_penalizing_function}(\gls{smoothing_matrix}(\gls{smoothing_parameter}))\frac{1}{\gls{sample_points}} \sum_{m=1}^{\gls{sample_points}} \left(x(t_m) - \estimated{x}(t_m; \gls{smoothing_parameter})\right)^2,
\end{equation}
where \(\gls{smoothing_penalizing_function}\) is a penalty function.
By default, the penalty is
\begin{equation}\label{eq:smothing_penalty_function}
\gls{smoothing_penalizing_function}(\gls{smoothing_matrix}(\gls{smoothing_parameter})) = \frac{1}{(1 - \tr(\gls{smoothing_matrix}(\gls{smoothing_parameter}))/\gls{sample_points})^2}.
\end{equation}
Additional penalty functions, such as \gls{akaike}, implemented as \fct{akaike}, or Shibata’s model selector, implemented as \fct{shibata}, are provided as well. 

In the code that follows, the smoothing functionality provided by \pkg{scikit-fda} is illustrated using the functions of the \emph{Phoneme} dataset \citep{hastie++_2009_elements}, which are rather noisy.

\begin{CodeInput}
import skfda
from skfda.preprocessing.smoothing import (
    kernel_smoothers,
    validation,
)

X, y = skfda.datasets.fetch_phoneme(return_X_y=True)

grid = validation.SmoothingParameterSearch(
    kernel_smoothers.KNeighborsSmoother(),
    [2, 3, 4, 5],
    scoring=validation.LinearSmootherGeneralizedCVScorer(validation.shibata),
)

grid.fit(X)
X_smooth = grid.transform(X)
\end{CodeInput}

\begin{figure}
  \includegraphics[width=\linewidth]{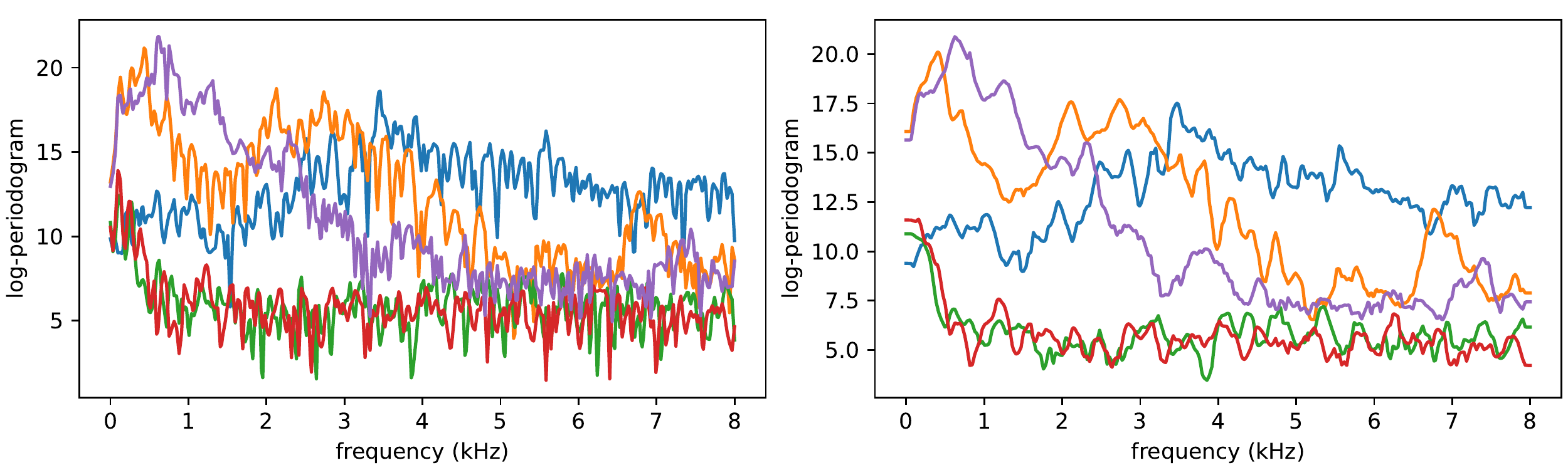}
  \caption[Kernel smoothing]{Five trajectories of the \emph{Phoneme} dataset before (left) and after (right) nearest neighbors smoothing. }
  \label{fig:smoothing}
\end{figure}

In this example, the methods of the class \class{KNeighborsSmoother} are used to smooth the data.
The optimal number of neighbors is selected between the values 2, 3, 4 and 5.
Finally, the \class{LinearSmootherGeneralizedCVScorer} with the \fct{shibata} penalty function is used for model selection.
The first five original curves and the smoothed ones are displayed in Figure~\ref{fig:smoothing}.

\subsubsection{Registration} 

Another type of preprocessing, which is especially relevant in FDA, is registration.
Registration consists in applying transformations to the raw data so that the functional observations are properly aligned.
There is a variety of reasons why misalignment can occur.
In some cases, it is the result of errors in the measurement process. 
In others, the domain has to be warped because the functions depend on an internal parameter, which is different from the one observed. 
For periodic functions, such as the signal of a heartbeat, the starting time for the different measurements could be different.
A number of strategies can be used for registration. 
For instance, maxima, minima, zeros, and other landmarks can be used as reference points for alignment. 
Alternatively, some measure of dispersion between the observations can be minimized. 
It is also possible to register a set of functional observations to a reference function.
After registration, it may be necessary to evaluate the functional observations at points in the domain that are different from the ones in the original grid.
This can be made utilizing the interpolation and extrapolation techniques described in Section~\ref{sec:additionalrepr}.
To carry out such an alignment, the package \pkg{scikit-fda} offers support for shift registration, and for elastic registration.

{
\newcommand{\samplefunctionn}{\varfunction{x}_i}
\newcommand{\transformedfunction}{\tilde{\varfunction{x}}}
\newcommand{\transformedfunctionn}{\transformedfunction_i}
\newcommand{\shiftn}{\gls{shift}_i}
\newcommand{\landmarktimen}{\tau_{i}}
\newcommand{\nlandmark}{P}
\newcommand{\landmarktimenfirst}{\tau_{i_1}}
\newcommand{\landmarktimenlast}{\tau_{i_\nlandmark}}
\newcommand{\landmarktimenl}{\tau_{il}}
\newcommand{\landmarktimex}{\tau^*}
\newcommand{\landmarktimexl}{\tau_l^*}
\newcommand{\warpingfunctionn}{\gls{warping_function}_i}

Shift registration consists in aligning the functional observations by a translation
\begin{equation}\label{eq:shift_registration}
	\transformedfunctionn(t) = \samplefunctionn(t + \shiftn), \quad i = 1, \ldots, \gls{nsamples},
\end{equation}
where \(\shiftn\) is the time shift applied to \(\samplefunctionn(t)\), and \(\transformedfunctionn(t)\) is the registered function \citep{ramsay+silverman_2005_functional}.
Shifting modifies the lower and upper bounds of the interval on which the function observations are defined. 
The values of the shifted functions that lie outside the original interval are discarded. 
For the subinterval in which the function values are not available, they are estimated by extrapolation. 
The method for extrapolation can be provided as an input. 

The shifting constants \(\left\{ \shiftn \right\}_{i=1}^{\gls{nsamples}} \) can be determined using different procedures. 
If a single landmark, such as a maximum, a minimum, or a zero crossing, is present in every curve, and their locations, \(\left\{ \landmarktimen \right\}_{i=1}^{\gls{nsamples}} \), are known, then the $i$-th curve can be shifted by \(\shiftn = \landmarktimen - \landmarktimex\). 
After registration, the location of the landmark is \(\landmarktimex\) for every curve, 
\begin{equation}\label{eq:shift_registration_landmarks}
	\transformedfunctionn(\landmarktimex) = \samplefunctionn(\landmarktimen), \quad i = 1, \ldots, \gls{nsamples}.
\end{equation} 
In \pkg{scikit-fda}, the function \fct{landmark\_shift\_registration} can be used to carry out this transformation.
The values of the \(\shiftn\) can be retrieved using the \fct{landmark\_shift\_deltas} function.

Alternatively, the values \(\shiftn\) can be computed by minimizing a least squares criterion \citep{ramsay+silverman_2005_functional}
\begin{equation}\label{eq:shift_registration_least_squares}
	\glsxtrshort{regsse} = \sum_{i=1}^{\gls{nsamples}} \int_{\gls{time_interval}}
        [\transformedfunction_i(t) - \estimated{\gls{mean_function}}(t)]^2 dt,
\end{equation}
where \(\estimated{\gls{mean_function}}(t)\) is the sample mean of the registered data \(\{\transformedfunction_i(t)\}_{i=1}^{\gls{nsamples}}\).
This type of shift registration can be performed with methods of class \class{LeastSquaresShiftRegistration}. 
Instead of the sample mean, which is the default value, a user-defined template function can be employed. 
In this case, the values for the \(\shiftn\) are stored as the attribute \code{deltas\_} after the registration.

Another type of registration available in the package \pkg{scikit-fda} is elastic registration. 
In elastic registration, one attempts to align the data by applying a warping transformation
\begin{equation}\label{eq:elastic_registration}
	\transformedfunctionn(t) = \samplefunctionn(\warpingfunctionn(t)), \quad i = 1, \ldots, \gls{nsamples}.
\end{equation}
The warping \(\warpingfunctionn\) is a monotonically increasing function defined in \(\gls{time_interval} = [a, b]\).
Assuming that the values of the function at the endpoints of this interval are fixed, it obeys the constraints \(\warpingfunctionn(a) = a\) and \(\warpingfunctionn(b) = b\).
If the locations of some landmarks are known, the warping function for elastic registration can be approximated by monotonically increasing splines \citep{ramsay+silverman_2005_functional}. 
Besides the boundary constraints specified earlier, the spline interpolator for the $i$-th functional observation has to satisfy
\begin{equation}\label{eq:elastic_registration_landmarks}
    \transformedfunctionn(\landmarktimexl) = \samplefunctionn(\landmarktimenl), \quad  l = 1, \ldots, L,
\end{equation}
where \(\left\{ \landmarktimenl \right\}_{l=1}^L\) are the landmark locations in the $i$-th observation, and \(  \landmarktimexl =  \warpingfunctionn^{-1}(\landmarktimenl)\) is the location of the \(l\)-th landmark in the registered curves. 
In \pkg{scikit-fda} this type of registration can be carried out using the function \fct{landmark\_elastic\_registration}. The warpings can be retrieved with the function \fct{landmark\_elastic\_registration\_warping}. 
A drawback of this approach is that the landmarks and their locations need to be identified beforehand.

An alternative type of elastic registration is to align the observations to a reference template. 
This has the advantage that no information on landmarks is needed.
In the elastic registration method described in \citet{srivastava++_2011_registration},
the template is defined in terms of the Karcher mean under the Fisher-Rao metric \citep{srivastava+klassen_2016_functional}. 
Then, an energy function depending on the Fisher-Rao distance between each curve and the template is minimized. 
To efficiently compute this distance, the square root velocity function (SRVF) transform is introduced \citep{joshi++_2007_novel}. 
The main reason for introducing this transform is that the Fisher–Rao distance between two functions is given by the $L^2$ distance between their SRVF representations.
The \pkg{scikit-fda}'s class \class{FisherRaoElasticRegistration} includes methods for this particular type of registration. 
The warping functions used are stored in the \code{warping\_} attribute of this class.
The implementation makes use of the dynamic programming routines for alignment to a template from the \proglang{Python} package \pkg{fdasrsf} \citep{tucker_2020_fdasrsf_python}.
This type of elastic registration is illustrated in Figure~\ref{fig:elastic_registration} with synthetic data. 
The original trajectories, which are displayed on the left plot, exhibit two local maxima whose relative locations are different in each of the curves.
In consequence, alignment cannot be achieved via a simple shift.
Note that after this type of elastic registration curves are well aligned even without previous information about the location of the landmarks. 

\begin{figure}
  \includegraphics[width=\linewidth]{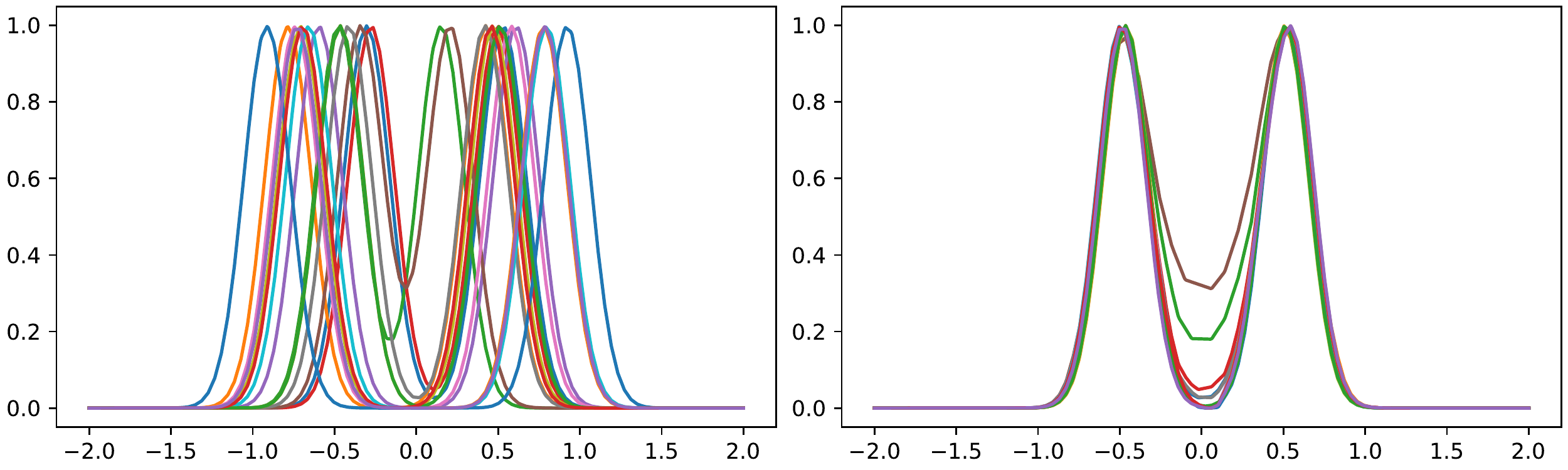}
  \caption[Elastic Fisher-Rao registration]{Elastic Fisher-Rao registration with synthetic data: The original curves are shown in the left panel. The registered curves are displayed in the right panel.}
  \label{fig:elastic_registration}
\end{figure}
}

\begin{figure}
  \includegraphics[width=\linewidth]{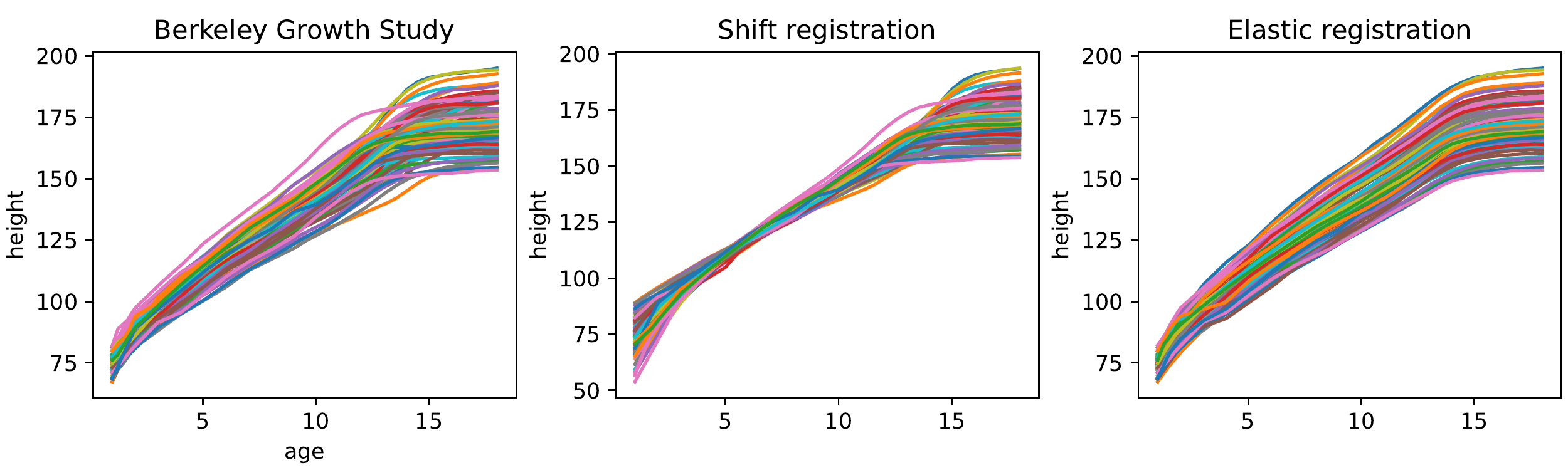}
  \caption[Comparison of registration methods]{Registration of the \emph{Berkeley Growth Study} data with different methods.
  From left to right: the original curves, shift registration by least squares and 
 elastic Fisher-Rao registration.
  }
  \label{fig:registration_comparison}
\end{figure}

The results of applying shift registration by least squares and elastic Fisher-Rao registration to the \emph{Berkeley Growth Study} data \citep{tuddenham_1954_physical} are compared in Figure~\ref{fig:registration_comparison}.
This figure has been generated using the following code:

\begin{CodeInput}
import skfda
from skfda.preprocessing.registration import (
    ElasticRegistration,
    ShiftRegistration,
)

X, y = skfda.datasets.fetch_growth(return_X_y=True)

X_aligned_elastic = ElasticRegistration().fit_transform(X)
X_aligned_shift = ShiftRegistration().fit_transform(X)

X.plot()
X_aligned_shift.plot()
X_aligned_elastic.plot()
\end{CodeInput}

The curves displayed in the left panel of Figure~\ref{fig:registration_comparison} trace the evolution of the heights of \(54\) girls and \(39\) boys since their birth until their \(18\)th birthday.
The nominal ages at which the measurements are made coincide for all individuals. 
However, each child has a different growth profile.
In particular, landmark features manifest themselves at different ages. 
For instance, even though most curves exhibit a growth spurt at puberty, the precise ages at which this occurs are different for each individual. 
Therefore, an elastic deformation of the actual age axis may uncover an internal age, which is more meaningful from a biological perspective.
The effects of shift and elastic registration based on the Fisher-Rao distance are displayed in the middle and right panels of Figure~\ref{fig:registration_comparison} respectively. 

\subsubsection{Dimensionality reduction}

Functional data are infinite-dimensional objects. 
Even in the case that they are represented by a set of discrete measurements, their dimensionality is typically very high.
In addition, nearby observations exhibit a large degree of dependence. 
Due to these characteristics, technical and computational difficulties arise in the analysis of these types of data.
To alleviate such difficulties one can represent the functional data in a lower-dimensional space while preserving as much information as possible.
The use of dimensionality reduction methods leads to gains in efficiency and, in some cases, improvements in interpretability and predictive capacity.
Furthermore, in this lower-dimensional representation, the methods of multivariate statistics can be employed \citep{vieu_2018_dimension}.

A simple dimensionality-reduction method is to select a set of impact or design points that are relevant for the task at hand; for instance, the most informative points for clustering, classification, or regression \citep{delaigle++_2012_componentwise,ferraty++_2010_mostpredictive,kneip++_2016_functional}.
Specifically, \pkg{scikit-fda}'s class \class{EvaluationTransformer} can be used to evaluate the functions as a set of points in the domain of the function.
Alternatively, a truncated basis representation can be used \citep{biau++_2005_functional, poskitt+sengarapillai_2013_description}. 
The coefficients of a functional data object represented as a basis expansion can be extracted using the methods of the class \class{CoefficientsTransformer}.
Besides these, the \pkg{scikit-fda} package provides methods for functional principal components analysis (FPCA) and variable selection methods.
These types of methods are described in what follows.

\paragraph{\Glsfmtlong{fpca}.}
Functional principal component analysis (FPCA) is a widely used dimensionality reduction method in FDA.
In this method, the individual functions are represented in the orthonormal basis of eigenfunctions of the stochastic process' covariance operator.
Dimensionality reduction is achieved by retaining the projections of the original functions onto the subspace of \(\gls{L2}\) spanned by the set of eigenfunctions that correspond to the largest eigenvalues. 
This representation is the one, among those of the same dimension, that explains the most of the data's variance.  

A random function $X \in \gls{L2}$ can be represented as
\begin{equation}\label{eq:fpca_basis}
    \randfunction{X}(t) = \gls{mean_function}(t) + \sum_{j=1}^{\infty} \xi_j\varFunction{\phi}_j(t),
\end{equation}
where \(\gls{mean_function}(t) = \mathbb{E}\left[\randfunction{X}(t)\right]\), \(\varFunction{\phi}_j(t)\) is the \(j\)-th principal component, and \(\xi_j = \int_{\gls{time_interval}} (\randfunction{X}(t)-\gls{mean_function}(t))\varFunction{\phi}_j(t)dt\), 
denotes the projection (score) along the \(j\)-th principal component. 
By the Karhunen-Loève Theorem, the scores \(\left\{\xi_j \right\}_{j\ge 1}\) are uncorrelated random variables \citep{wang++_2016_functional}.
Smoothed versions of the principal components can also be computed by applying the regularization penalties described in Section~\ref{sec:regularization}, using the procedure described in Section 9.4.2 of \cite{ramsay+silverman_2005_functional}.
The smooth principal components are obtained by optimizing a function that takes into account not only the sample variance but also a term that penalizes the roughness of the principal components.
A reduction of the dimension of the data can be achieved by truncating the basis expansion in Equation~\ref{eq:fpca_basis}, so that only the first $K$ components are included 
\begin{equation}\label{eq:fpca_basis_truncated}
    \randfunction{X}(t) = \gls{mean_function}(t) + \sum_{j=1}^{K} \xi_j\varFunction{\phi}_j(t). 
\end{equation}

In \pkg{scikit-fda}, functional principal component analysis can be carried out using the methods of class \class{FPCA}.
The following code illustrates this functionality for the \emph{Berkeley Growth Study} data:

\begin{CodeInput}
import skfda
import matplotlib.pyplot as plt

X, y = skfda.datasets.fetch_growth(return_X_y=True)

fpca = skfda.preprocessing.dim_reduction.feature_extraction.FPCA(
    n_components=2,
)
fpca.fit(X)

skfda.exploratory.visualization.fpca.FPCAPlot(
    X.mean(), fpca.components_, multiple=30,
).plot()

scores = fpca.transform(X)
scores_class_0 = scores[y == 0]
scores_class_1 = scores[y == 1]

plt.figure()
plt.scatter(scores_class_0[:, 0], scores_class_0[:, 1])
plt.scatter(scores_class_1[:, 0], scores_class_1[:, 1])
\end{CodeInput}

\begin{figure}
  \includegraphics[width=\linewidth]{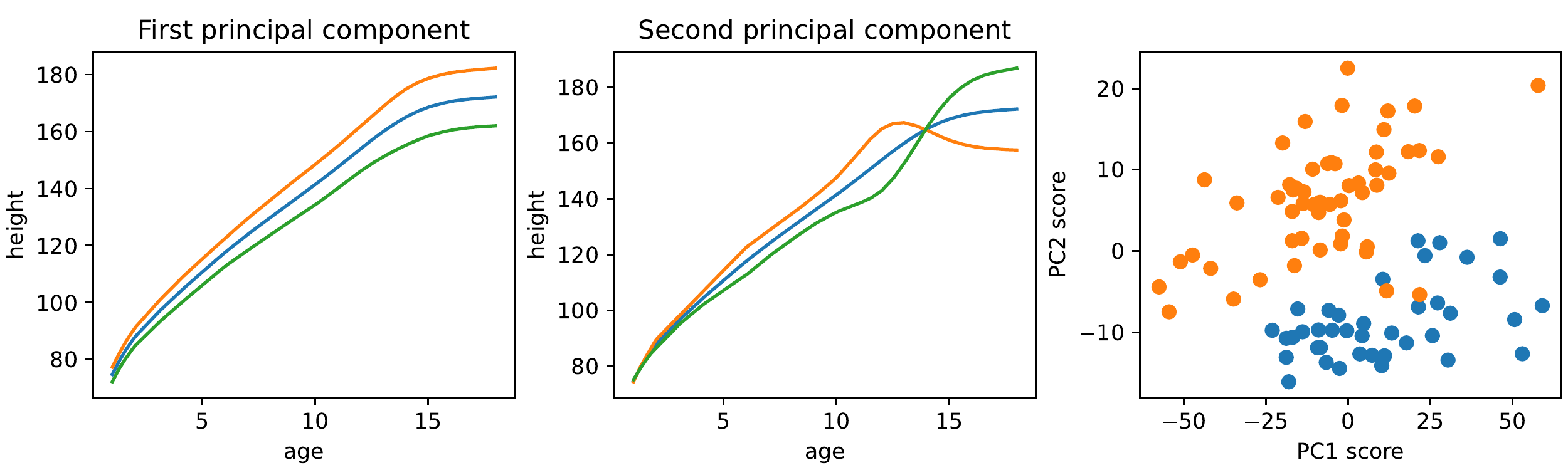}
  \caption{Principal components analysis for the \emph{Berkeley Growth Study} data. 
  The first and second principal components are plotted in the leftmost panel as perturbations around the mean (in blue). 
  In the right panel, the scores of individual functional observations for the first two components are plotted. 
  The orange and blue points correspond to girls' and boys' growth curves, respectively.}
  \label{fig:fpca_components}
\end{figure}

In this example, the first two principal components are computed. 
Then, the functional observations are projected onto the two-dimensional subspace spanned by these components.
A numerical quadrature is used to compute the corresponding inner products.
The class \class{FPCAPlot} is used to display the curves \(\left\{\mu(t)\pm\phi_j(t); \ j = 1, 2\right\}\), which are the result of adding and subtracting the $j$-th eigenfunction to the sample mean. 
In the case of the Berkely growth study, the first component captures overall deviations (either positive or negative) with respect to the mean. 
The second one reveals patterns associated to differences of growth speed. 
In particular, it exhibits a maximum followed by a sign change at around puberty. 
The resulting plots are shown in the left and middle panels in Figure~\ref{fig:fpca_components}. 
Finally, the projection of the curves onto the first two principal components is obtained using the \attribute{transform} method of the class \class{FPCA}.
The scores of these two components are displayed as points in the right panel of Figure~\ref{fig:fpca_components}.
Note that, with some exceptions, boys (blue) and girls (orange) appear grouped in two separate clusters in this plot.

\paragraph{Variable selection.}

The package \pkg{scikit-fda} provides a variety of tools to carry out variable selection.
A simple approach is to apply a multivariate variable selection method to the discretized representation of the functional observations  \citep{berrendero++_2016_mrmr,jimenez+maldonado_2021_automatic}.
The \pkg{scikit-fda} class \class{EvaluationTransformer} can be used to transform \class{FData} objects into \pkg{NumPy} arrays.
Then, any \proglang{Python} library for multivariate variable selection can be used. 
If this approach does not take into account the functional nature of the data, there can be difficulties in the analysis \citep{aneiros+vieu_2016_sparse}.
A multivariate method that takes into account the redundancy that arises from the continuity of functional data is \gls{mrmr} \citep{ding+peng_2005_minimum, peng++_2005_feature, berrendero++_2016_mrmr}. 
This method is implemented in \pkg{scikit-fda} in the class \class{MinimumRedundancyMaximumRelevance}. 
The dependence measures that quantify the relevance and the redundancy can be specified by the user.

In addition, the \pkg{scikit-fda} package includes a collection of variable selection methods that specifically take into account the functional nature of the data: a method based on the theory of Reproducing Kernel Hilbert Spaces (RKHS-VS), maxima hunting (MH), and recursive maxima hunting (RMH). 

The RKHS-VS method, implemented in the class \class{RKHSVariableSelection}, was introduced for binary classification problems \citep{berrendero++_2018_use}. 
For a specified value of $d$, the goal is to identify the set  of points \(\varvector{t} = (t_1, \ldots, t_d)^\top \in \mathcal{T}^d\) and select the corresponding function values, $\randfunction{X}(t_1), \ldots, \randfunction{X}(t_d)$, that maximize the Mahalanobis distance between groups
\begin{equation}\label{eq:rkhs_vs}
	(\mu_1(\varvector{t})-\mu_0(\varvector{t}))^\top K(\varvector{t}, \varvector{t})^{-1} (\mu_1(\varvector{t})-\mu_0(\varvector{t})),
\end{equation} 
where \(\mu_0(\varvector{t}), \mu_1(\varvector{t})\), and K(\varvector{t}, \varvector{t}) are the mean functions of each class and the covariance function evaluated at  $\randfunction{X}(t_1), \ldots, \randfunction{X}(t_d)$, respectively.
In homoscedastic binary classification problems, with a fixed dimension $d$ this selection is optimal in terms of classification error.
In practice, the exploration of all possible combinations is often infeasible. To reduce the computational costs, a greedy search is implemented.

In MH \citep{berrendero++_2016_variable, ordonez++_2018_determining} one selects the values of \(t \in \gls{time_interval}\) that correspond to local maxima of a non-negative dependence measure between \(\randfunction{X}(t)\) and the class label. 
The selected variables are thus the most relevant in a region.
Furthermore, the values of \(\randfunction{X}(t)\) that are close to these local maxima, which generally provide redundant information, are automatically discarded.
In \cite{berrendero++_2016_variable}, the distance correlation \citep{szekely++_2007_measuring} is used as the dependence measure. 
This variable selection method is implemented in the class \class{MaximaHunting}, using 
the implementation of distance correlation function provided by the \pkg{dcor} package \citep{ramos_2020_dcor}.
MH is an interpretable, fully functional method with optimal performance in an important class of functional classification problems.
In spite of its simplicity and good performance, MH has some limitations.
Specifically, there can be numerical difficulties to identify the local maxima of the depence measure.
Futhermore, MH takes into account only the marginal relevance of a single variable.
Variables that are only relevant when selected in combination with other cannot be identified by these procedures. 

RMH \citep{torrecilla+suarez_2016_feature} addresses these limitations by assuming a particular form of the stochastic process from which the trajectories are sampled.
The algorithm proceeds as follows: First the value of \(t\) that is the global maximum of the dependence between the variable \(\randfunction{X}(t)\) and the class label is selected. 
Let \(t^*\) be such optimum and, therefore, \(\randfunction{X}(t^*)\) the variable selected. 
The information conveyed by \(\randfunction{X}(t^*)\) is removed by subtracting from the trajectories the conditional expectation of the process given the value of the selected variable. 
Then, the global maximum of the resulting process is identified.
The algorithm proceeds in this iterative manner until a pre-specified number of variables have been selected, or until a convergence criterion is fulfilled.
In \pkg{scikit-fda}, the class \class{RecursiveMaximaHunting} provides an enhanced, very customizable implementation of this method.

\subsection{Exploratory analysis}

Exploratory analysis methods are used to identify salient features, visualize, and describe the data from a statistical point of view.
Specifically, the \pkg{scikit-fda} package provides tools for the computation of summary statistics, including robust ones, interactive tools for visual analysis, and outlier detection. 

\subsubsection{Summary statistics}

Common summary statistics, such as the sample mean function and the sample covariance function can be estimated using the tools provided by \pkg{scikit-fda}.
Consider a set of functional observations \(\left\{ \varfunction{x}_i(t) \right\}_{i=1}^{\gls{nsamples}}\).
The sample mean,
\begin{equation}\label{eq:mean}
	\estimated{\gls{mean_function}}(t) = \frac{1}{\gls{nsamples}}\sum_{i=1}^{\gls{nsamples}} \varfunction{x}_{i}(t),
\end{equation}
can be computed by applying the function \fct{mean} to the \class{FData} object in which the data are stored.
The functional observations can be either in discrete form or in a basis representation.
The resulting mean function is a \class{FData} object of the same type as the input (i.e., discretized or in a basis representation).

The sample covariance function \(\estimated{\gls{covariance_function}}\),
\begin{equation}\label{eq:covariance}
	\estimated{\gls{covariance_function}}(t, s) = \frac{1}{\gls{nsamples} - 1}\sum_{i=1}^{\gls{nsamples}} (\varfunction{x}_{i}(t) - \estimated{\gls{mean_function}}(t))(\varfunction{x}_{i}(s) - \estimated{\gls{mean_function}}(s)),
\end{equation}
can be computed applying the function \fct{cov} to the corresponding \class{FData} object.
Similarly, the function \fct{var} can be used to computed the sample variance, \(\estimated{\gls{covariance_function}}(t, t)\).
Irrespective of the representation of the functional observations, the sample variance and covariance are returned in discretized form.
Instead of the functions \fct{mean}, \fct{cov} and \fct{var}, the \class{FData} methods of the same name can be used to compute these summary statistics.

\begin{figure}[t]
  \centering
  \includegraphics[width=\linewidth]{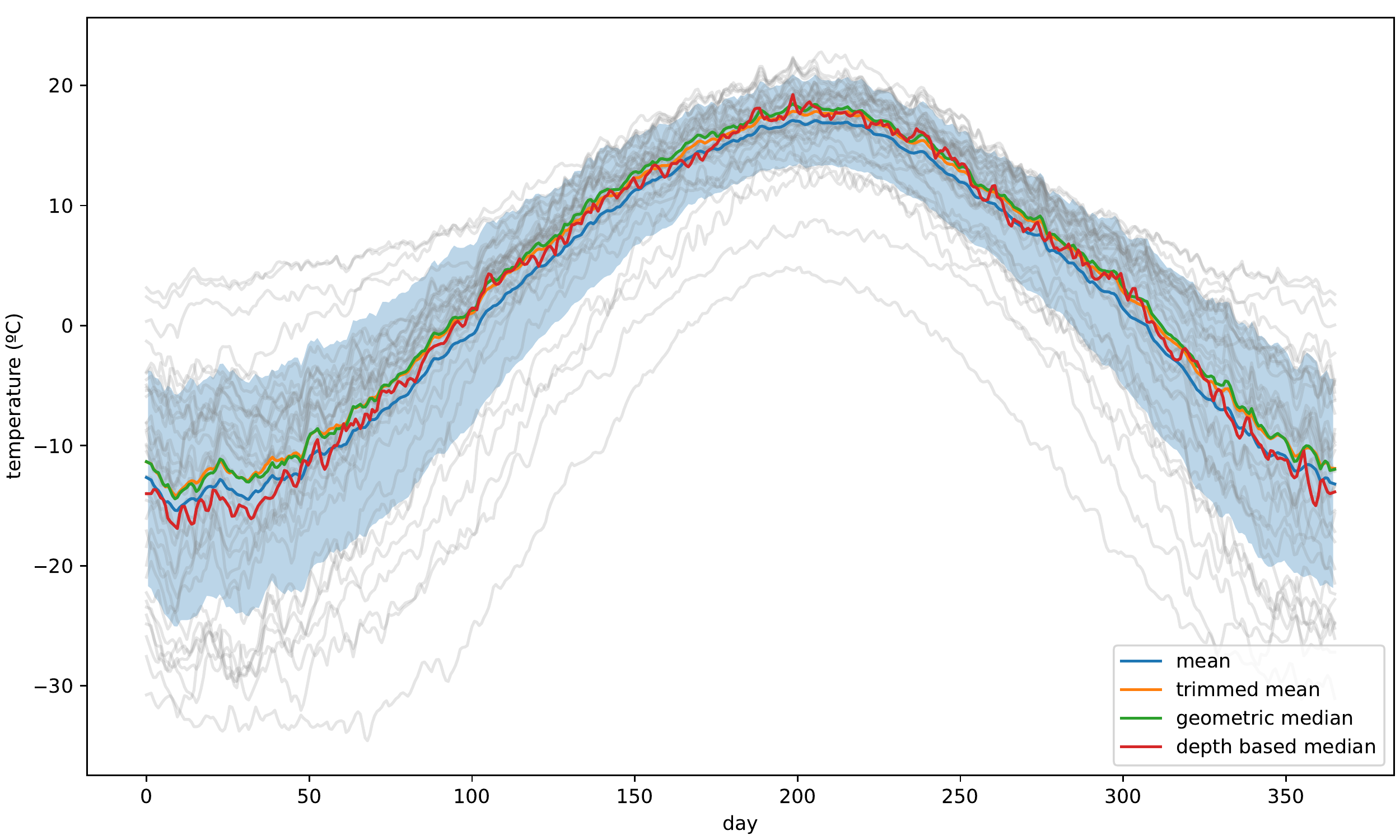}
  \caption[Centrality statistics]{Centrality statistics of the \emph{Canadian Weather} dataset. The shaded band corresponds to one standard deviation around the mean.}
  \label{fig:stats}
\end{figure}

Figure~\ref{fig:stats} presents an illustration of this functionality for the \emph{Canadian Weather} dataset. 
The sample estimate of the mean temperature curves is shown as a blue curve.
The shaded area corresponds to one standard deviation around the estimated mean.
Other measures of centrality, such as the trimmed mean, the geometric mean, and the median, are displayed in this figure as well. 
These robust statistics will be described in some detail in Section~\ref{sec:robust}.

\subsubsection{Depth measures}\label{sec:depth}

Depth measures quantify the centrality of a function in relation to a set of functions. 
These measures are used for exploratory analysis, to compute robust statistics, detect outliers, and for data visualization (e.g., the functional box-plot).
In contrast to the univariate case, a variety of definitions of functional depth can be given that yield different orderings of the functional observations in the sample.
Each of these functional depths lead to different definitions of robust statistics and of degrees of outlyingness. 

Some common functional depth measures are implemented in the package \pkg{scikit-fda}.
In particular, the methods of class \class{IntegratedDepth} can be used to compute integrated depth measures \citep{fraiman+muniz_2001_trimmed}, which are averages of univariate depths. 
Specifically, the integrated depth of the function \(\varfunction{x}\) is 
\begin{equation} \label{eq:integrated_depth_measure}
	\text{ID}(\varfunction{x}) = \int_{\gls{time_interval}} \gls{depth_function}(\varfunction{x}(t))dt.
\end{equation}
where \(\gls{depth_function}\) is an univariate depth function.
This function can be selected by the user.  
The default is the measure proposed by \cite{fraiman+muniz_2001_trimmed}:
\begin{equation}\label{eq:fraiman_muniz_univariate_depth}
	\gls{depth_function}(x(t)) = 1 - \left\lvert \frac{1}{2}- F_{X(t)}(x(t))\right\rvert,
\end{equation}
where \(F_{X(t)}\) denotes the distribution function of the marginal.

An alternative definition is the \gls{bd}, introduced by \cite{lopez+romo_2009_concept}. 
To compute this functional depth, one needs to identify the bands that are delimited by all possible pairs of functional observations in the sample.
The BD value is the fraction of bands that completely encompass the curve.
In \pkg{scikit-fda}, this quantity can be computed using methods of the class \class{BandDepth}.
A related, less restrictive measure, is the \gls{mbd}.
This measure takes into account not only the number of bands that contain \(x\), but also the time that \(x\) lies within each band. 
The MBD has better statistical properties than the original BD, in part because it is an integrated depth measure \citep{nagy++_2016_integrated}.
In \pkg{scikit-fda}, MBD is implemented in the class \class{ModifiedBandDepth}.

\subsubsection{Robust statistics}\label{sec:robust}

The package \pkg{scikit-fda} provides support for the computation of robust statistics. 
Robust statistics may provide a better characterization of the data than non-robust ones (e.g., the mean or the covariance functions), especially in the presence of outliers. 
One of the most important robust statistics is the geometric median \citep{lardin++_2014_analysing}
\begin{equation}\label{eq:geometric_median}
	\median = \underset{\varfunction{z} \in \mathcal{F}}{\arg \min} \sum_{i=1}^{\gls{nsamples}} \left \| \varfunction{x}_i-\varfunction{z} \right \|.
\end{equation}
It can be computed with the function \fct{geometric\_median}.
Alternatively, the median can be defined as the deepest point in the sample.
Different depth measures yield different definitions of the median.
These types of medians can be computed with the function \fct{depth\_based\_median}. 

Functional depth measures can be used also to define the degree of outlyingness of a function in a sample: the larger the depth value the more central the functional observation is. 
Finally, funcional depth measures can be used to define trimmed means \citep{fraiman+muniz_2001_trimmed}.
A trimmed mean is a robust version of the standard mean in which the most outlying functional observations (the ones with the lowest depth values) are discarded. 
In \pkg{scikit-fda}, the trimmed mean is implemented in function \fct{trim\_mean}.

The geometric median, the MBD-based median, and the MBD-based trimmed mean in which 10\% of the data are discarded, of the \emph{Canadian Weather} dataset are shown in Figure~\ref{fig:stats}.

\subsubsection{Interactive visualization tools and outlier detection}

Visualization tools can be used to gain insight into the data. 
In particular, trends, salient features, and other patterns in the data can be identified simply by inspection.
Visualization tools can be utilized also to single out functional observations that are markedly different from the other observations in the sample (outliers). 
Outlier detection is useful to identify rare events, novel patterns, anomalies, or erroneous measurements. 
The package \pkg{scikit-fda} provides a number of interactive tools for data visualization and outlier detection. 
Their implementation utilizes the functionality provided by \pkg{matplotlib} \citep{hunter_2007_matplotlib}.

\begin{figure}
  \centering
  \includegraphics[width=\linewidth]{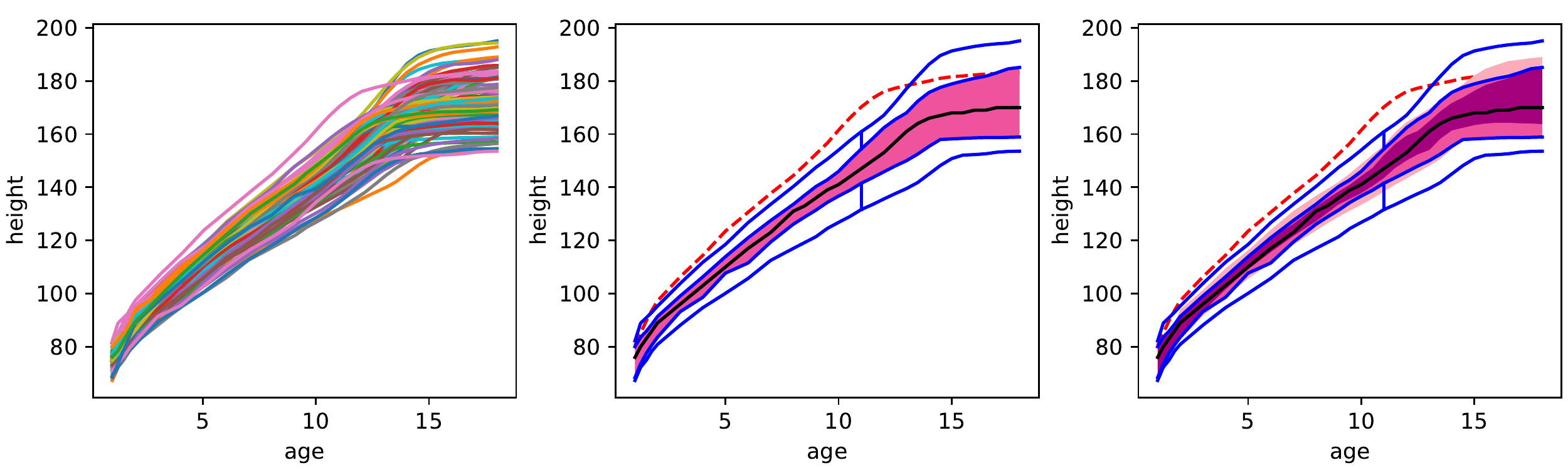}
  \caption[Functional boxplot]{Functional boxplots of the \emph{Berkeley Growth Study} dataset. 
  The original curves are depicted in the left panel.
  The standard functional boxplot is shown in the central panel. 
  In this panel, The black line stands for the functional median.
  The central envelope is displayed as a pink band around the median. 
  The blue whiskers and their fences mark the maximum non-outlying envelope. Outliers are shown as a red dashed lines. 
  In the right panel, different shades of pink are used for the deepest \(25\%\), \(50\%\), and \(75\%\)  of the data.
  }
  \label{fig:boxplot}
\end{figure}

Functional data objects have a \fct{plot} method that can be used to graph the curves. 
Some customization options, such as group colors or labels, are available for this method.  
An illustration of its use with the \emph{Berkeley Growth Study} dataset is shown on the left-hand panel of Figure~\ref{fig:boxplot}.
For \class{FDataGrid} objects, the \fct{scatter} method can be used to display the values of the function as individual points in a graph. 
This method was used to generate Figure~\ref{fig:notation}.

Another tool for visual exploration provided by \pkg{scikit-fda} is the functional boxplot \citep{sun+genton_2011_functional}.  
This is an generalization of the univariate boxplot for functional data. 
The functional boxplot consists of a graph of the functional median (i.e., the deepest curve in the sample) surrounded by a central envelope, which encompasses the deepest 50\% of the observations, and a maximum non-outlying envelope. 
The width of this outer envelope is determined by scaling the central one by a constant factor.
This constant factor can be selected by the user. 
Its default value is \(1.5\).
In \pkg{scikit-fda}, the class \class{Boxplot} can be used to generate and customize functional boxplots.
In this plot, a trajectory is marked as an outlier if it lies beyond the maximum non-outlying envelope for some interval.
The class \class{BoxplotOutlierDetector} can be used for outlier detection based on this criterion.
Some customizable elements of \class{Boxplot} objects are the depth measure, and the definition of centered bands that encompasses a user-specified fraction of the deepest observations. 
The following code provides an illustration of these functionalities with the \emph{Berkeley Growth Study} dataset. 
The plots that result from the execution of this code are displayed in Figure~\ref{fig:boxplot}.

\begin{CodeInput}
import skfda

X, _ = skfda.datasets.fetch_growth(return_X_y=True)

X.plot()

boxplot = skfda.exploratory.visualization.Boxplot(
    X,
    depth_method=skfda.exploratory.depth.ModifiedBandDepth(),
)
boxplot.plot()

boxplot = skfda.exploratory.visualization.Boxplot(
    X,
    depth_method=skfda.exploratory.depth.ModifiedBandDepth(),
    prob=[0.75, 0.5, 0.25],
)
boxplot.plot()
\end{CodeInput}

Another tool for functional data visualization and outlier detection is the \glsfirst{ms} \citep{dai+genton_2018_multivariate, dai+genton_2019_directional}.
In this method, the degree of outlyingness of a functional observation is characterized in terms of two quantities: the magnitude outlyingness (MO) and the shape outlyingness (VO). 
The MS-plot is the scatter plot of the values MO and VO for each functional observation.
This two-dimensional representation of the data can be used, for instance, to identify clusters of functions, or detect potential outliers, either in shape or in magnitude. 

\begin{figure}[t]
  \includegraphics[width=\linewidth]{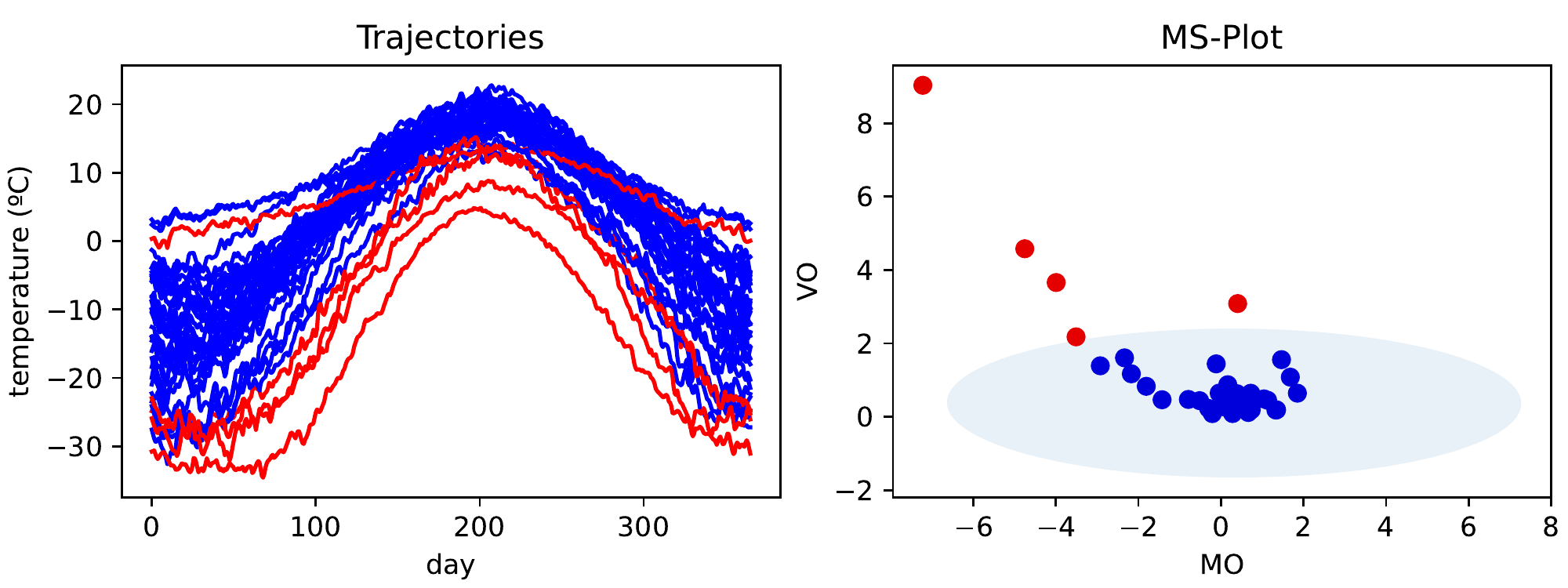}
  \caption{\gls{ms} and outliers of the \emph{Canadian Weather} dataset. The original yearly temperature curves are displayed in the left panel. The corresponding \gls{ms} is shown in the right panel. The observations identified as outliers by the \gls{ms} criterion (those outside the blue ellipse) are marked in red.}
  \label{fig:msplot}
\end{figure}

The following code can be used to display the \gls{ms} for the temperature curves of the \emph{Canadian Weather} dataset together with the original trajectories. Additionally, outliers are identified according to the MS-plot criterion and marked in red.
The class \class{MagnitudeShapePlot} generates the MS-plot and uses internally the methods of the class \class{MSPlotOutlierDetector} for outlier detection.
The resulting plots are shown in Figure~\ref{fig:msplot}. 

\begin{CodeInput}
import skfda

X, y = skfda.datasets.fetch_weather(return_X_y=True)
X = X.coordinates[0]

ms_plot = skfda.exploratory.visualization.MagnitudeShapePlot(X)
ms_plot.plot()

fig = X.plot(
    group=ms_plot.outliers,
    group_colors=["blue", "red"],
)
\end{CodeInput}

\begin{figure}[t]
  \centering
  \includegraphics[width=\linewidth]{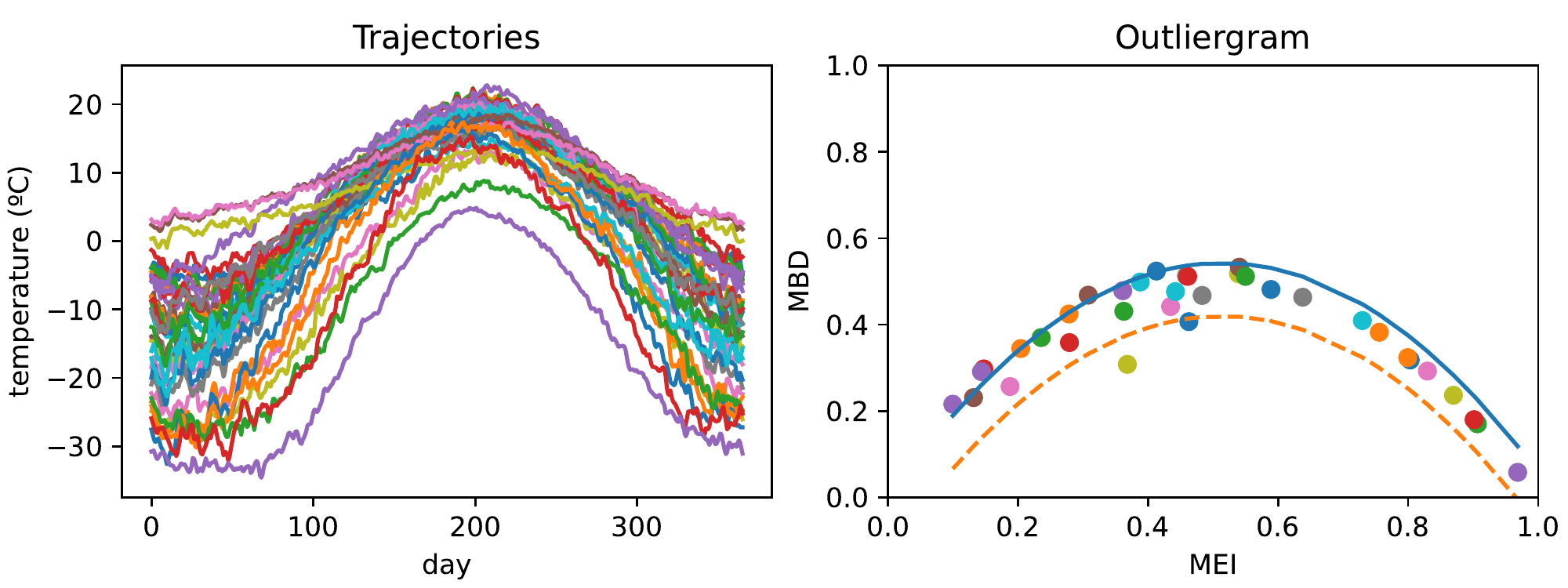}
  \caption[Outliergram]{Outliergram of the yearly temperature curves for the \emph{Canadian Weather} dataset. 
  The original trajectories are in the left panel and the corresponding outliergram is shown in the right panel. 
  The blue line corresponds to the reference parabola. 
  The orange dashed line separates the typical curves (above) from the outliers (below).}
  \label{fig:outliergram}
\end{figure}

The class \class{Outliergram} provides an additional method for data visualization and detection of shape outliers \citep{arribas+gil_2014_shape}.
The graph is defined in terms of two related quantities:  the \gls{mei} and the \gls{mbd}.
The MEI of a trajectory is the average over time of the fraction of curves in the sample that lie above it. 
Each curve is a point (MEI, MBD) in the scatter plot.
The outliergram takes advantage of the fact that points corresponding to typical functional observations lie on a parabola, whose analytical form is known.
This parabola is used as a reference for the identification of shape outliers.
Specifically, the degree of outlyingness of a curve is quantified in terms of its vertical distance to the parabola. 
The \pkg{scikit-fda}'s classes \class{Outliergram} and \class{OutliergramOutlierDetection} can be used to generate the outliergram and to detect outliers by using this criterion, respectively.
The following code illustrates this functionality with the temperatures of the \emph{Canadian Weather} dataset. The original trajectories and the corresponding outliergram are shown in Figure~\ref{fig:outliergram}.

\begin{CodeInput}
import skfda
import matplotlib.pyplot as plt

X, y = skfda.datasets.fetch_weather(return_X_y=True)
X = X.coordinates[0]

fig = X.plot()
fig = skfda.exploratory.visualization.Outliergram(X).plot()
\end{CodeInput}

In addition to standard plotting capabilities, most graphs generated with \pkg{scikit-fda} incorporate some interactive features.
For example, the cursor can be placed at a point in the graph to display the actual coordinate values and the label of the observation.
In addition, if different plots are used for visual exploration of some functional dataset, selecting a particular curve in one plot highlights the corresponding curve in the other active plots.
Finally, widgets such as sliders can be used to select curves by  some property, such as the label of the observation, or their depth in the sample.

\begin{figure}[t]
  \centering
  \includegraphics[width=\linewidth]{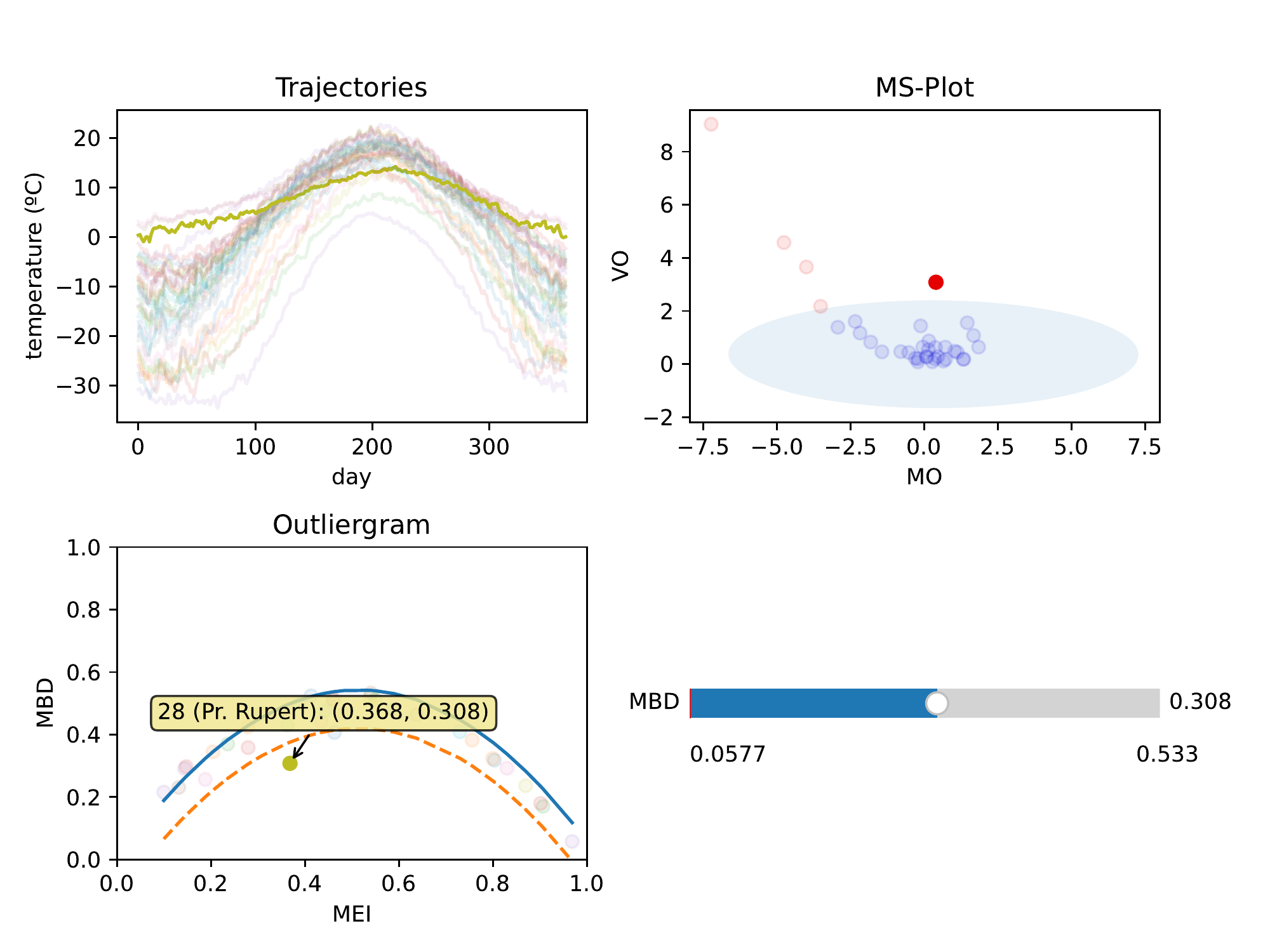}
  \caption[Interactive visualization]{Interactive features in multiple plots for the \emph{Canadian Weather} dataset.}
  \label{fig:interactive}
\end{figure}

An illustration of this interactive functionality is presented in Figure~\ref{fig:interactive}.
In this figure, three different kinds of plots are displayed for the temperature curves of the \emph{Canadian Weather} dataset: a graph of the sample trajectories, the \gls{ms}, and the outliergram.
In the lower right side a slider has been created that displays the \gls{mbd} value of the selected curve.
A functional observation can be selected either by choosing a value in the slider widget or by clicking on the corresponding point in the \gls{ms} or in the outliergram. 
The datum selected is then highlighted in all plots. 
Finally, the cursor has been placed at a point in the MS-plot.
This brings up a tooltip in which relevant information of the corresponding functional observation is displayed.

\subsection[Integration with scikit-learn for machine learning]{Integration with \pkg{scikit-learn} for machine learning}

The \pkg{scikit-fda} package has been especially designed for seamless integration with \pkg{scikit-learn} \citep{pedregosa_2011_scikit-learn}.
Specifically, there are a number of methods that transform the functional data into a two-dimensional array so that \pkg{scikit-learn}'s machine learning algorithms can be applied.
For instance, the method \fct{transform} of the class \class{EvaluationTransformer} returns an array with the values of the functions in the sample at a specified set of points.
The coefficients of a functional data object represented as a basis expansion can be extracted using the methods of the class \class{CoefficientsTransformer}.
Additionally, other \pkg{scikit-fda} methods, such as variable selection, can be used to this end.

The provided classes and methods for preprocessing conform to \pkg{scikit-learn}'s \gls{api} \citep{buitinck_2013_api}.
An advantage of adopting this standard is that they can be employed in a \pkg{scikit-learn} pipeline (class \class{Pipeline}). 
The following code illustrates how to build such a pipeline for a classification problem with functional data:
\begin{CodeChunk}
% Extracted from code/sklearn_example.py
\begin{CodeInput}
import skfda

from sklearn.model_selection import GridSearchCV, train_test_split
from sklearn.pipeline import Pipeline
from sklearn.svm import SVC

import skfda.preprocessing.smoothing as smoothing
import skfda.preprocessing.dim_reduction as dimred

X, y = skfda.datasets.fetch_phoneme(return_X_y=True)

X_train, X_test, y_train, y_test = train_test_split(
    X, y, random_state=0)

smoothing_step = smoothing.kernel_smoothers.KNeighborsSmoother()
dimred_step = dimred.feature_extraction.FPCA()
classification_step = SVC()

pipeline = Pipeline([
    ('smoothing', smoothing_step),
    ('dimred', dimred_step),
    ('classification', classification_step)])

grid = GridSearchCV(
    pipeline,
    param_grid={
        'smoothing__smoothing_parameter': [3, 5, 7],
        'dimred__n_components': [1, 2, 3],
        'classification__C': [0.001, 0.01, 0.1, 1, 10],
    })

grid.fit(X_train, y_train)

score = grid.score(X_test, y_test)

print(f"{score:.3}")
\end{CodeInput}
% Extracted from code/stdout/sklearn_example
\begin{CodeOutput}
0.879
\end{CodeOutput}
\end{CodeChunk}

In this pipeline, the following sequence of operations is applied: nearest-neighbors kernel smoothing, dimensionality reduction by functional PCA, and, finally, a standard (multivariate) \gls{svm} classifier with an RBF kernel. 
The hyperparameters of the whole model, including those that correspond to preprocessing, are tuned by cross-validation using \pkg{scikit-learn}'s class \class{GridSearchCV}.
Specifically, the number of neighbors in the nearest-neighbors kernel smoother is selected between the values \(3\), \(5\), and \(7\).
The number of principal components considered ranges from \(1\) to \(3\).
The grid \([0.001, 0.01, 0.1, 1.0, 10.0]\) is explored to determine the regularization parameter of the \gls{svm}. 
Then, the best values of the hyperparameters are used to fit the model using the complete training set. 
Finally, the accuracy of the classifier is computed and printed. 

\section{Code quality and documentation} \label{sec:documentation}

The \pkg{scikit-fda} library has been designed to provide a powerful, self-contained, flexible, stable, and easy-to-use framework for the analysis of functional data in \proglang{Python}.
The package is built as a SciPy Toolkit (SciKit)\footnote{See \url{https://svn.scipy.org/scikits.html} for further details on SciKits (Accessed 2021-12-23)}.
It is fully integrated in the \proglang{Scientific Python} software ecosystem\footnote{\url{https://scientific-python.org/} (Accessed 2021-12-23)}.
\proglang{Scientific Python} is a collection of free and open-source \proglang{Python} packages for scientific and technical computing \citep{oliphant_2007_python, millman+aivazis_2011_python}.
Standard coding and naming practices  are used throughout the project \citep{vanrossum++_2001_style, goodger+vanrossum_2001_docstring}. 
This not only improves the legibility of the code and simplifies its maintenance, but also facilitates external contributions to the development of the package.
Whenever appropriate, the design conforms to \pkg{scikit-learn} specifications \citep{pedregosa_2011_scikit-learn}, so that the machine learning tools implemented in that package can be readily applied to functional data.
To ensure the quality and robustness of the software, a comprehensive suite of unit and integration tests is provided. 
These automated tests are executed regularly in a continuous integration environment.
We have attempted also to minimize the number of dependencies and to provide flexible interfaces that are intuitive and consistent throughout the application. 

The \pkg{scikit-fda} package is free and open-source software distributed under the OSI-approved 3-Clause BSD license\footnote{\url{https://opensource.org/licenses/BSD-3-Clause} (Accessed 2021-12-23)}.
The version described in this paper is 0.7.1 and was released on 2022-01-25.
The GitHub page \url{https://github.com/GAA-UAM/scikit-fda} is the main communication channel with the developers of the package for questions, bug reports, and feature requests.
Contributions from the members of the FDA community are encouraged, as are comments and suggestions to improve the quality of the software.

\begin{figure}
  \centering
  \hfill
  \includegraphics[width=0.40\linewidth]{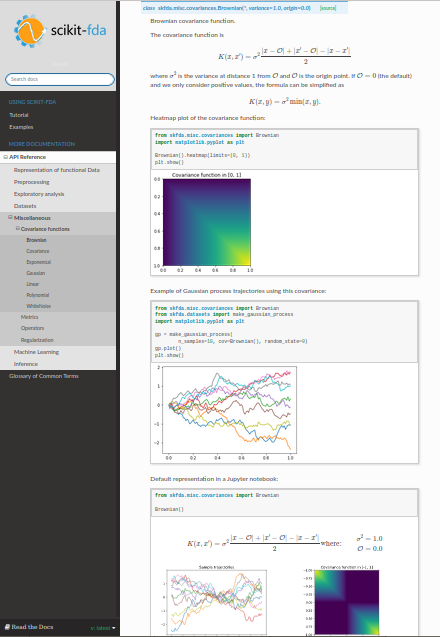}
  \hfill
  \includegraphics[width=0.40\linewidth]{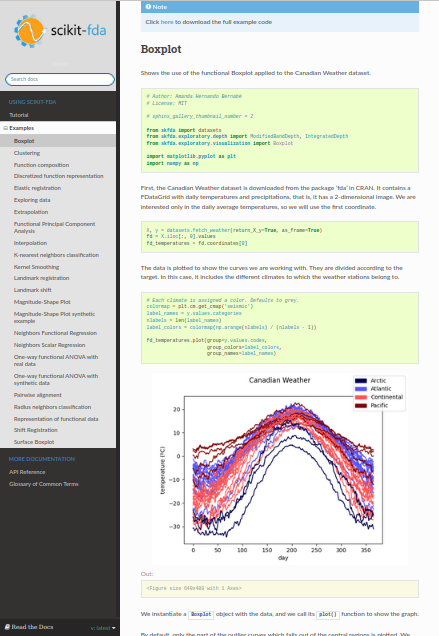}
  \hfill
  \caption[Documentation of the package]{\pkg{scikit-fda} documentation: On the left panel, the reference page of the Brownian covariance function is shown. On the right one, an illustration of the functional boxplot functionality.}
  \label{fig:docs}
\end{figure}

To facilitate the use of \pkg{scikit-fda}, exhaustive documentation, including installation instructions, tutorials, API references, and illustrative examples are provided. 
The documentation, which is available online at \url{https://fda.readthedocs.io}, is built with the \proglang{Python} tool \pkg{Sphinx} \citep{sphinx_2020_sphinx}.
The examples and tutorials have been devised with \pkg{Sphinx-Gallery} \citep{najera++_2020_sphinx}.
They can be viewed online or downloaded as interactive \pkg{Jupyter} notebooks \citep{kluyver++_2016_jupyter}. 
In Figure~\ref{fig:docs}, two screenshots of the documentation pages are shown:
A reference page for the Brownian covariance function and an example of use 
of the functional boxplot functionality are displayed on the left and right panels of the figure, respectively.

\section*{Acknowledgments}

The authors want to express their gratitude to the developers who made contributions to the \pkg{scikit-fda} package.
In particular, we thank David García Fernández, Amanda Hernando Bernabé, Yujian Hong, Pedro Martín Rodríguez-Ponga Eyriès, Pablo Pérez Manso, Elena Petrunina, Luis Alberto Rodríguez Ramírez, and Álvaro Sánchez Romero for their participation in the project.
The authors acknowledge financial support from the Spanish Ministry of Education and Innovation, projects PID2019-106827GB-I00 /
AEI / 10.13039/501100011033 and PID2019-109387GB-I00.
This research was also supported by an FPU grant (Formación de Profesorado Universitario) from the Spanish Ministry of Science, Innovation and Universities(MICIU) with reference FPU18/00047.

\bibliographystyle{apalike}

\bibliography{refs}

\newpage

\end{document}